\documentclass[12pt]{article}
 \pdfoutput=1
 
 \usepackage[colorlinks=true,
 linkcolor=red,
 urlcolor=blue,
 citecolor=blue]{hyperref}
\usepackage[makeroom]{cancel}
\usepackage[normalem]{ulem}
\usepackage{amsmath,amssymb}
\usepackage{cleveref}
\usepackage{xcolor}
\usepackage{graphicx}
\usepackage{cite}
\usepackage{pdflscape}
\usepackage{multirow}
\usepackage[thinlines]{easytable}
\usepackage{url}
\usepackage[utf8]{inputenc}
\usepackage{longtable}
\usepackage{booktabs}
\usepackage{mathrsfs}
\usepackage{subcaption}
\usepackage{float}
\usepackage{pstricks}
\usepackage[toc,page]{appendix}
\usepackage[mathscr]{euscript}

\usepackage{color}

\newcommand{\eref}[1]{Eq.~\eqref{eq:#1}}

\newcommand{\tref}[1]{Table~\ref{tab:#1}}

\newcommand{\nnl}{\nonumber \\}

\newcommand{\beq}{\begin{equation}} 
\newcommand{\eeq}{\end{equation}} 
\newcommand{\ba}{\begin{array}}  
\newcommand{\ea}{\end{array}} 
\newcommand{\bea}{\begin{eqnarray}}  
\newcommand{\eea}{\end{eqnarray} }  
\newcommand{\be}{\begin{eqnarray}}  
\newcommand{\ee}{\end{eqnarray} }  
\newcommand{\bal}{\begin{align}}
\newcommand{\eal}{\end{align}}   
\newcommand{\bi}{\begin{itemize}}  
\newcommand{\ei}{\end{itemize}}  
\newcommand{\ben}{\begin{enumerate}}  
\newcommand{\een}{\end{enumerate}}  
\newcommand{\bc}{\begin{center}}
\newcommand{\ec}{\end{center}} 
\newcommand{\bt}{\begin{table}}
\newcommand{\et}{\end{table}}  
\newcommand{\btb}{\begin{tabular}}
\newcommand{\etb}{\end{tabular}}

\newcommand{\gev}{\mathrm{GeV}}

\def\q2 {q^2}

\def\bt{\begin{table}}
\def\et{\end{table}}

\def\mET{E_T \hspace{-1.0em}/\;\:}

\begin{document}

\vspace*{-2cm}
\begin{flushright}
LPT Orsay 19-32
\end{flushright}

\begin{center}
\vspace{1.2cm}
{\LARGE \bf Associated $Z^\prime$ production \\ 
in the flavorful $U(1)$ scenario for $R_{K^{(*)}}$}
\vspace{1.4cm}

\renewcommand{\thefootnote}{\fnsymbol{footnote}}
{\sf ~Siddharth Dwivedi$^{a,}$\footnote{tpsd5@iacs.res.in},
 Adam Falkowski$^{b,}$\footnote{adam.falkowski@th.u-psud.fr}, Dilip Kumar Ghosh$^{a,}$\footnote{tpdkg@iacs.res.in},  
and Nivedita Ghosh$^{a,}$\footnote{tpng@iacs.res.in}}
\renewcommand{\thefootnote}{\arabic{footnote}}
\setcounter{footnote}{0}

\vspace*{.8cm}
\centerline{\it $^a$ School of Physical Sciences, Indian Association for the Cultivation of Science} 
\centerline{\it 2A $\&$ 2B, Raja S.C. Mullick Road, Kolkata 700032, India}   
\centerline{${}^b$\it Laboratoire de Physique Th\'{e}orique (UMR8627), CNRS, Univ. Paris-Sud,}
\centerline{\it Universit\'{e} Paris-Saclay, 91405 Orsay, France}\vspace{1.3mm}

\vspace*{.2cm}

\end{center}

\vspace*{10mm}
\begin{abstract}\noindent\normalsize
The flavorful $Z^\prime$ model with its couplings restricted to the left-handed second generation leptons
and third generation quarks can potentially resolve the observed anomalies in $R_K$ and $R_{K^*}$. After examining 
the current limits on this model from various low-energy processes, we probe this scenario at 14 TeV high-luminosity 
run of the LHC using two complementary channels: one governed by the coupling of $Z'$  to $b$-quarks and the other to
muons. We also discuss the implications of the latest LHC  high mass resonance searches in the dimuon channel on the 
model parameter space of our interest. 
\end{abstract}

\newpage 

\renewcommand{\theequation}{\arabic{section}.\arabic{equation}} 

\section{Introduction}
\setcounter{equation}{0}

In the last few years, the LHCb collaboration has reported a number of deviations from $\mu$-$e$ universality in B-meson processes. In particular, the ratios of $\mu^+ \mu^-$ to $e^+ e^-$ final states in $B \to K^{(*)} \ell^+ \ell^-$ decays:  $R_K$~\cite{Aaij:2019wad} and $R_{K^*}$~\cite{Aaij:2017vbb} are observed to be smaller than one, each displaying a $\sim 2.5\sigma$ deviation from lepton-flavor universality predicted by the Standard Model (SM). 
Recent global analyses~\cite{Alguero:2019ptt,Ciuchini:2019usw,Aebischer:2019mlg,Kowalska:2019ley,Arbey:2019duh}, which also take into account other $b\rightarrow s \ell^+ \ell^-$ mediated processes, conclude that the SM is disfavored by the current experimental data with a confidence level exceeding $5\sigma$.      

The global fit can be significantly improved if the effective Lagrangian below the weak scale contains new contributions to the 4-fermion operator $(\bar b_L \gamma^\rho s_L) (\bar \mu_L \gamma_\rho \mu_L)$, in addition to the ones  generated by the exchange of SM particles in loops. 
One option to arrange for these contributions is to assume that the high-energy theory contains a new electrically neutral vector particle $Z'$ coupled to muons and, in a flavor-violating way, to bottom and strange quarks. 
In this scenario, the 4-fermion operator in question can arise from tree-level $Z'$ exchange. 
There is already a vast literature discussing $Z'$ models explaining the  $b\to s \ell \ell$ anomalies, see e.g.~\cite{Gauld:2013qba,Buras:2013qja,Altmannshofer:2014cfa,Crivellin:2015mga,
Crivellin:2015lwa,Niehoff:2015bfa,Celis:2015ara,Greljo:2015mma,Niehoff:2015iaa,Altmannshofer:2015mqa,Falkowski:2015zwa,
Carmona:2015ena,GarciaGarcia:2016nvr,Megias:2016bde,Chiang:2016qov,Altmannshofer:2016oaq,Boucenna:2016qad,Foldenauer:2016rpi,
Kamenik:2017tnu,Chivukula:2017qsi,Faisel:2017glo,Ellis:2017nrp,Alonso:2017uky,Carmona:2017fsn,Dalchenko:2017shg,Raby:2017igl,
Bian:2017rpg,Bian:2017xzg,Alok:2017jgr,Falkowski:2018dsl,Kohda:2018xbc,Fox:2018ldq,Chala:2018igk,Darme:2018hqg,Allanach:2018odd,Ko:2019tts,Biswas:2019twf,
Allanach:2019mfl,Alok:2019ufo,Kawamura:2019rth}.
A generic feature of these model is that the $Z'$ is within the kinematic reach of the LHC and thus can be searched for directly. 
In particular,  these models always predict a non-zero cross section for the quark-level process $b(\bar b) \bar s (s) \to Z' \to \mu \mu$, which leads to the dimuon resonance signature at the LHC. 
Furthermore, in some models the $Z'$ coupling to $b s$ is correlated with couplings to other quarks, which opens further production channels at the LHC~\cite{Chivukula:2017qsi,Dalchenko:2017shg}. 

The goal of this paper is to study new LHC signatures of the $Z'$ boson responsible for the  $b\to s \ell \ell$ anomalies. 
We consider the model described in Ref.~\cite{Falkowski:2018dsl} where $Z'$, in addition to the coupling to muons, also possesses a sizable coupling to $b \bar b$. 
This model predicts several new signatures where $Z'$ is produced in association with some SM particles.  
We focus on two such signatures, which we find especially promising:  
\begin{itemize}
 \item $ p p \to Z^\prime +  1b (2b) \to \mu^+\mu^-  +  1b (2b)$, 
 \item $p p \to Z' \mu^{\pm} + \cancel{E}_T \to 3 \mu + \cancel{E}_T$.
\end{itemize}
For these two processes we study the discovery prospects at the LHC run 3 and the subsequent high-luminosity phase (HL-LHC). 
We show that the above signature can be observed with the significance exceeding $5 \sigma$ in the parameter space of the $Z'$ model favored by the $b\to s \ell \ell$ anomalies and consistent with all other experimental constraints. 
The information obtained by studying these two processes is complementary to that conveyed by generic dimuon resonance searches, and will be crucial for the identification of the microscopic model responsible for the $b\to s \ell \ell$ anomalies.


In what follows, in Section \ref{model} we discuss the model and list the range of couplings of the $Z'$ to muons and b-quarks allowed by low-energy precision measurements. 
In Section~\ref{sec:Analysis} we present a detailed analysis of LHC prospects of discovering the $Z'$ in two complementary channels where the $Z'$ is produced in association with SM particles. The production rate of $Z'$ in the two channels is  governed by its coupling either to $b$-quarks or to muons and thus they can potentially probe different regions of the allowed parameter space dominated by either of the two couplings. 
In Section~\ref{sec:dimuon} we compare the sensitivity of these associated $Z'$ production searches with that of the generic dimuon resonance searches.. 
Finally, we summarise and conclude in Section \ref{Summary}.

\section{The model}
\label{model}
\setcounter{equation}{0}

 We consider a massive spin-1 boson $Z'$ with coupling to quarks and leptons that can address the $R_K$ and $R_K^*$ anomalies. 
We work with the setup described in Ref.~\cite{Falkowski:2018dsl}, however in this paper we assume that only the $Z'$ boson can be produced at the energy scale available at the LHC.    
The relevant BSM interactions pertaining to our collider analysis are encoded in the following Lagrangian:
\begin{equation}
{\cal L} \supset Z'_\mu \left(
g_{bb} \bar{q}_{L} \gamma^\mu q_{L} 
+ g_{bs}\bar{b}_{L} \gamma^\mu s_{L} 
+ g_{\mu \mu}\bar L_{L} \gamma^\mu L_{L} 
\right),
\label{eq:Zp_Rk_couplings}
\end{equation}
where $q_L = (t_L,b_L)^T$, $L_L = (\nu_{\mu \, L}, \mu_L)^T$.
The  $Z'$  couplings $g_{\mu\mu}$, $g_{bb}$, and $g_{bs}$ to muons, $s$- and $b$-quarks are in principle free parameters. However, in the setup of~\cite{Falkowski:2018dsl} in the absence of fine-tuning one expects $|g_{bs}| \sim  |V_{ts}g_{bb}|$, where $|V_{ts}|\approx 0.04$ is the 3-2 entry of the CKM matrix.  
In the following for simplicity we assume $g_{bs} = V_{ts} g_{bb}$, and that $g_{bb}$ and $g_{\mu \mu}$ have the same sign. 
Thus, the parameter space in our analysis is 3-dimensional, and consists of the 2 couplings $g_{bb}$, $g_{\mu \mu}$ and the $Z'$ mass $M_{Z'}$.  

Integrating out the $Z'$ boson generates four-fermion contact interactions in the effective theory below the scale $M_{Z'}$. 
In particular, a new contribution to the effective interaction  $(\bar{b}_{L} \gamma_\rho s_{L})(\bar \mu_{L} \gamma^\rho \mu_{L})$ is generated, adding to the SM  contribution induced at the loop level.  This is the scenario with $C_{9\mu}^{\rm NP} = - C_{10\mu}^{\rm NP}$, using  the standard notation of flavor physics. 
Such a pattern of new physics corrections provides a very good fit to the  measured $R_K$, $R_K^*$, and other $b \to s \mu \mu$ observables~\cite{Alguero:2019ptt,Ciuchini:2019usw,Aebischer:2019mlg,Kowalska:2019ley,Arbey:2019duh}.  The best fit of Ref.~\cite{Aebischer:2019mlg},  $C_{9\mu}^{\rm NP} = - C_{10\mu}^{\rm NP} = -0.53 \pm 0.09$, translates into the following constraint on our parameters:
\begin{equation}
\label{eq:bestfit}
{g_{bb} g_{\mu \mu} \over  M_{Z'}^2 }    = { 1.00 \pm 0.17 \over (6.9~\text{ TeV})^2} \qquad @~68\%~{\rm CL}. 
\end{equation}
In the following of this analysis we will assume that the values of the parameters correspond to this best fit within $1\sigma$ uncertainty.   

There are further low-energy constraints on these parameters. 
One is due to four-lepton interactions generated by integrating out $Z'$, which are constrained by the  trident muon production in neutrino scattering~\cite{Geiregat:1990gz,Mishra:1991bv,Altmannshofer:2014pba}. 
Using the global fit of Ref.~\cite{Falkowski:2017pss} one finds 
\begin{equation}
 {g_{\mu \mu}^2 \over  M_{Z'}^2}  \lesssim  {1 \over (330~\text{GeV})^2}  \qquad @~99\%~{\rm CL}. 
\end{equation}
Another combination of the model parameters is probed thanks to the $Z'$ generating the $\Delta F=2$ operator $(\bar b_L \gamma_\mu s_L)^2$,  which affects the $B_s$ meson mass difference $\Delta m_{B_s}$.  
In some of the previous literature this contribution is constrained by comparing the experimentally measured $\Delta m_{B_s}$ with the one predicted by the SM. 
This is however not quite correct. 
The reason is that the SM prediction $\Delta m_{B_s}^{\rm SM}$  is a function of the CKM parameters, which are obtained from a global fit to flavor observables (see e.g. \cite{Charles:2004jd}). 
 {\em These fits  always include  the $B_s$ meson mass difference  as one of the inputs}. 
As a result, the CKM parameters and consequently the predicted   $\Delta m_{B_s}^{\rm SM}$    can be ``contaminated" by the new physics contribution of the $Z'$, 
and it is not consistent to use  the $\Delta m_{B_s}$ observable alone to constrain $Z'$. 
Instead, consistent constraints can be obtained by comparing  {\em  different} observables that probe the same CKM parameters but are affected {\em  differently} by the $Z'$. 
Such an analysis was performed in Ref.~\cite{Descotes-Genon:2018foz} which compared CKM parameters extracted from  $\Delta m_{B_s}$ with those extracted from $B \to D^{(*)} \ell \nu$  decays. 
That analysis leads to the constraint 
\begin{equation}
 {g_{bb}^2 \over  M_{Z'}^2}  \lesssim  {1 \over (11.5~\text{TeV})^2}  \qquad @~99\%~{\rm CL}. 
\end{equation}

\begin{figure}[!t]
\begin{center}
\includegraphics[width=0.55 \textwidth]{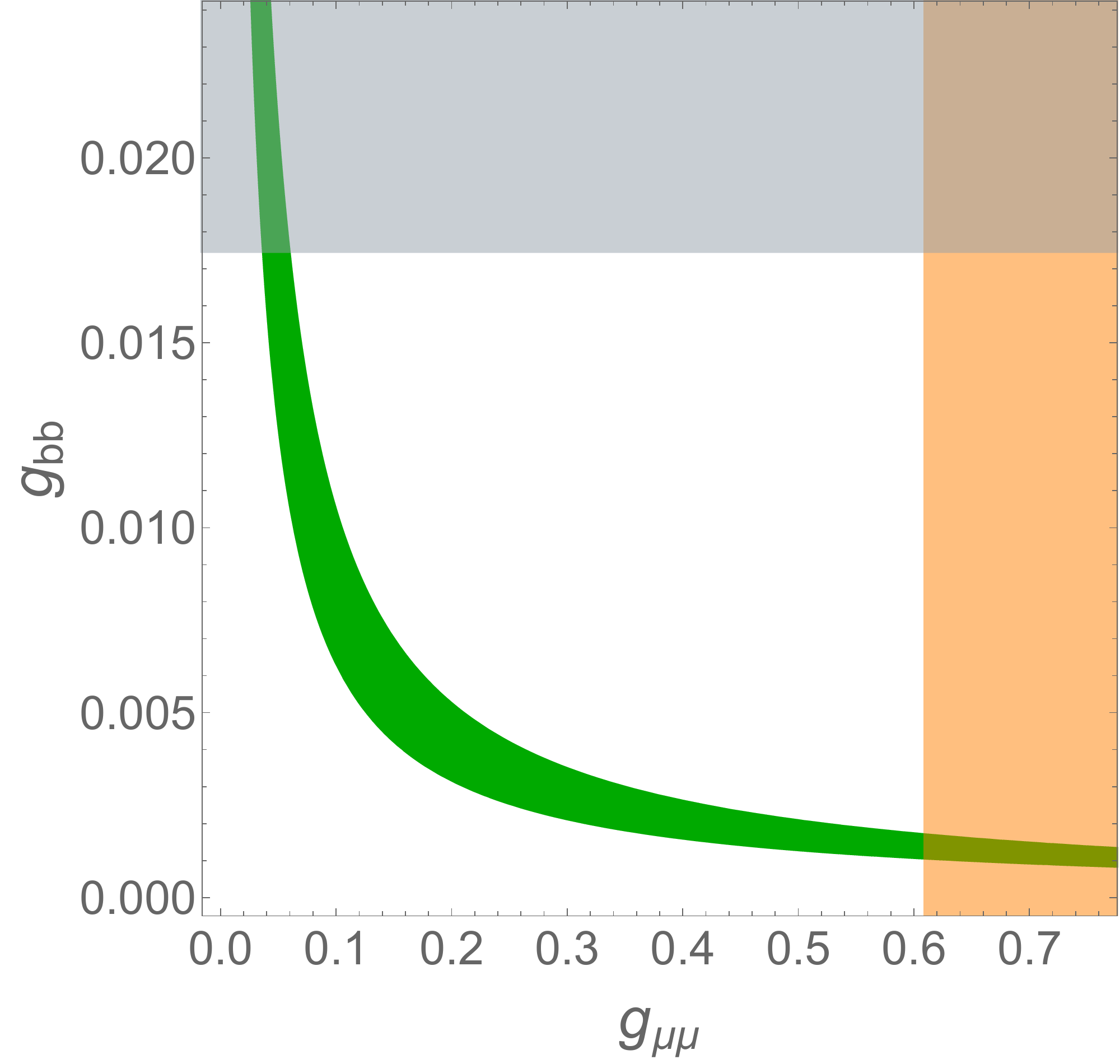}
\caption{
The parameter space in the $(g_{\mu\mu},g_{bb})$ plane for $M_{Z'} = 200~\gev$ preferred at 68\% CL by the $b\to s\ell^+\ell^-$  anomalies (parabolic green band). 
We also show   the regions excluded at  99\% CL. by trident neutrino production 
(vertical orange band), and by the analysis  Ref.~\cite{Descotes-Genon:2018foz} comparing the $\Delta m_{B_s}$ and $B \to D^{*} \ell \nu$ probes of the CKM elements (horizontal grey band).  
} 
\label{fig:parspace}
\end{center}
\end{figure}

An example of the parameter space is shown in Figure \ref{fig:parspace} for $M_{Z'} = 200$~GeV. 
Clearly, fitting the $b \to s \mu \mu$ anomalies together with the  low-energy constraints discussed above leaves a finite interval for the $Z'$ coupling $g_{\mu \mu}$ and $g_{bb}$. 
The intervals $g_{\mu \mu}^{\rm min} \lesssim g_{\mu \mu} \lesssim g_{\mu \mu}^{\rm max}$ and  $g_{b b}^{\rm min} \lesssim g_{b b} \lesssim g_{b b}^{\rm max}$ allowed at 99\% CL for the particular values of $M_{Z'}$ used in our collider analysis are shown in Table~\ref{benchmark1}. 

\begin{table}[h!]
\centering
\begin{tabular}{|c|c|c|c|c|c|}
\hline
$M_Z'$ (GeV) & $g^{\rm min }_{\mu\mu}$ &  $g^{\rm max }_{\mu\mu}$ & $g_{\mu \mu}^{1 \sigma}$ & $g^{\rm min }_{bb}$ & $g^{\rm max }_{bb}$ \\ \hline
200          & 0.040 & 0.61 & [0.067,0.078]  & 0.0016 &  0.017   \\ \hline
300          & 0.060 & 0.91 & [0.10,0.12]  & 0.0024 &  0.026   \\ \hline
500          & 0.10 & 1.5  & [0.16,0.20] &  0.0040 & 0.044    \\ \hline
750          & 0.15	& 2.3 & [0.24,0.32] &	0.0060 & 0.065 \\ \hline
1000	     & 0.20	& 3.0  & [0.32,0.43] & 0.0080 &   0.087   \\ \hline
\end{tabular}
\caption{Intervals for the couplings  $g_{\mu\mu}$ and $g_{bb}$ consistent with explaining the $b \to s \ell \ell$ anomalies, and not excluded at  99\% CL by the CKM~\cite{Descotes-Genon:2018foz}  and trident~\cite{Altmannshofer:2014pba} constraints. 
We also show the 1$\sigma$ confidence interval for the coupling $g_{\mu \mu}$ obtained from the likelihood combining the above mentioned constraints. 
}
\label{benchmark1} 
\end{table}

\section{Collider Analysis} 
\label{sec:Analysis}
\setcounter{equation}{0}

In this section we discuss LHC signatures of a $Z'$ boson with a pattern of couplings to matter motivated by the $b \to s \mu \mu$ anomalies, as given in \eref{Zp_Rk_couplings}.
One signature, already discussed in several previous works \cite{Dalchenko:2017shg,Falkowski:2018dsl}, is the resonant dimuon production, 
$pp \to Z' \to \mu^+ \mu^-$. 
In this scenario, the $Z^\prime$ is predominantly produced at the LHC via the $b \bar b$ fusion, with a subleading contribution from the $b \bar s$ and $\bar b s$ fusion, and it decays to a pair of muons with a branching fraction that is strongly dependent on the couplings $g_{bb}$ and $g_{\mu \mu}$.
Another  signature is $ pp \to Z \to 4 \mu$ \cite{Altmannshofer:2014cfa,Altmannshofer:2014pba,Altmannshofer:2016jzy}, where the $Z$ boson first decays to two muons, and then a $Z'$ (off-shell or on-shell, depending on its mass) is radiated off one of the muons.

The goal of this paper is to explore alternative signatures of the $Z'$ boson at the LHC. 
We focus on the following two processes: 
\begin{itemize}
    \item $p p \to Z^\prime +  1b (2b) \to \mu^+\mu^-  +  1b (2b)$, 
    \item $p p \to Z^\prime \mu^\pm \cancel{E_T} \to 3 \mu^\pm + \cancel{E_T}$. 
\end{itemize}

\begin{figure}[h]
	\centering
		\includegraphics[width=90mm,height=40mm]{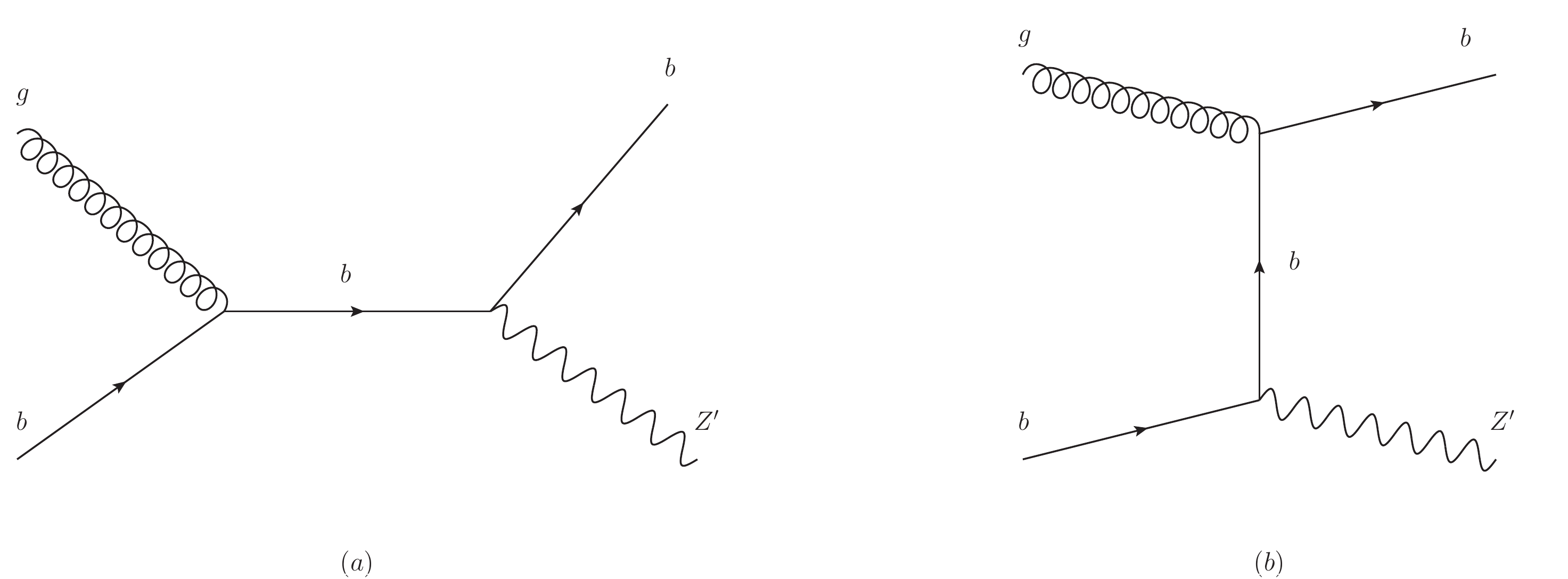} \\
		\includegraphics[width=90mm,height=60mm]{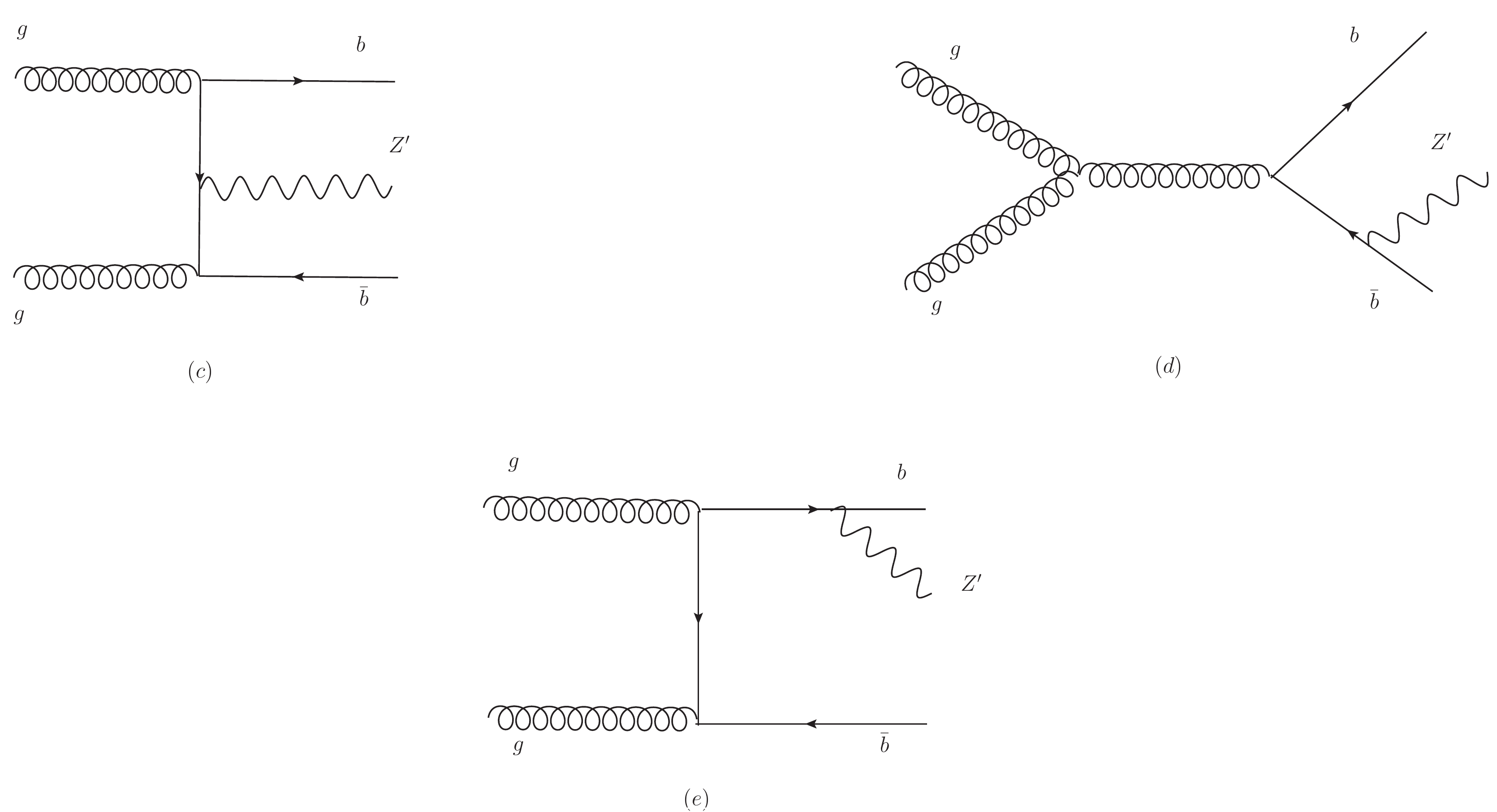} \\
	\caption{Leading Feynman diagrams for the $Z' + 1b(2b)$ final state.}
	\label{fig:Feynman_Diag_2mu1b2b} 
\end{figure}

\begin{figure}[h]
	\centering
			\includegraphics[width=90mm,height=35mm]{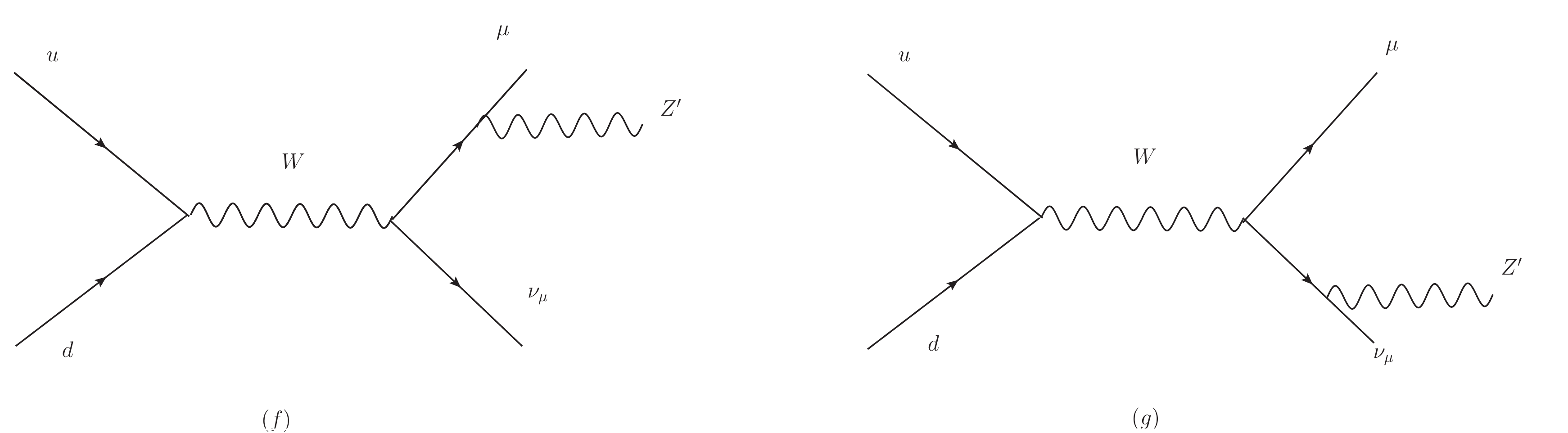}
	\caption{Leading Feynman diagrams for the $Z' \mu^{\pm} \cancel{E_T}$ final state.}
	\label{fig:Feynman_Diag_3mu_missET} 
\end{figure}

The leading Feynman diagrams for these processes are shown in Figure \ref{fig:Feynman_Diag_2mu1b2b} and \ref{fig:Feynman_Diag_3mu_missET}. 
In the first process the $Z'$ boson is radiated off a b-quark, while in the second it is radiated off a muon or a neutrino. 
In both cases we study the situation where the $Z'$ decays to a muon pair. 
Consequently, the rate of the first process depends on both $g_{b b}$ and $g_{\mu \mu} $ couplings, while in the second case it depends only on $g_{\mu \mu}$. 
Note that, following  Eq.~(\ref{eq:bestfit}), the magnitude of $g_{b b}$ and $g_{\mu \mu}$ is anti-correlated in our scenario. 
For this reason, the two processes target complementary regions of the parameter space: 
the $3 \mu^\pm + \cancel{E_T}$ signal is more relevant for larger $g_{\mu \mu}$, 
while the $\mu^+\mu^-+b$ signal is more relevant for smaller $g_{\mu \mu}$. 

We implemented the interactions in \eref{Zp_Rk_couplings} in {\tt FeynRules}\cite{Alloul:2013bka} so as to generate a MadGraph5 model file. 
We then generated both the signal as well as SM backgrounds events using $\texttt{MadGraph5\_aMC@NLO}$~\cite{Alwall:2014hca} 
at the leading order (LO) and at the  parton level.  
For the parton distribution function (PDF) we used the \texttt{NN23LO1} implementation~\cite{Ball:2014uwa}. 
The parton level events are passed to {\tt PYTHIA 8}~\cite{Sjostrand:2014zea} for showering and hadronization.
Finally, the showered events are passed through the detector level simulation using
{\tt Delphes3}~\cite{deFavereau:2013fsa},  with the jets reconstructed using the anti-$k_{T}$ jet algorithm \cite{Cacciari:2008gp}.
In our analysis we ignore $Z'$ production proceeding via the $Z^\prime$-$b$-$s$ coupling, 
which is suppressed due to the smallness of that coupling in our model, $g_{ bs}/g_{bb} \sim |V_{ts}| = {\cal O}(10^{-2})$.

\subsection{$ p p \to Z^\prime +  1b (2b) \to \mu^+\mu^-  +  1b  (2b)$ channel}
\label{sec:mumubb}
In this channel we consider the production of the $Z'$ boson in $\sqrt{s}= 14$~TeV LHC in association with either one or two $b$-quarks, 
followed by the $Z'$ decay into a muon pair. 
The dominant background contributions for this signal arise from the SM processes $p p \to  \mu^+ \mu^- + jets$,  $ p p \to t \bar t + jets \to b \bar b W^+ W^- + jets 
\to  b \bar b \mu^+ \mu^- \nu_{\mu} \bar \nu_{\mu}+ jets$. 
Here $jets$ denote both the light jets and $b$-jets. The light jets are taken into account since they 
can contribute to the background via being mistagged
as $b$-jets. For the $\mu^+ \mu^- + jets$ and for $t \bar t + jets$ background, events are matched up to three jets and two jets 
respectively by kt-MLM matching scheme \cite{Mangano:2006rw,Hoche:2006ph}.

\begin{figure}[h!]
	\centering
		\includegraphics[width=0.7\textwidth]{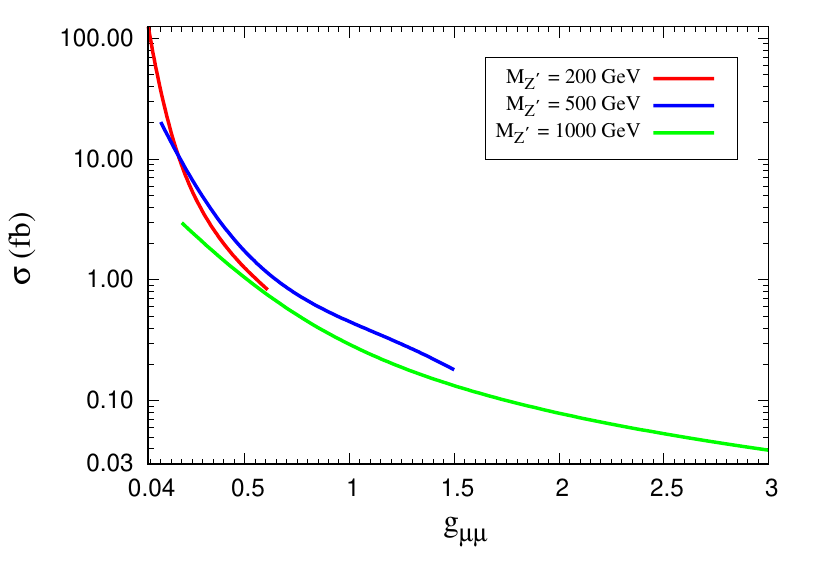} 
	
	\caption{
	The signal cross-section as a function of $g_{\mu\mu}$ for the $p p \to Z^\prime +  1b (2b) \to \mu^+\mu^-  +  1b  (2b)$
	process.
We show the results for $M_{Z'}=$200, 500~{\rm and}~1000 GeV at $\sqrt s=$14 TeV.
 Each curve is plotted for the corresponding $g_{\mu\mu}$ range taken from  Table~\ref{benchmark1}, 
which is determined by flavor and trident  constraints. 
 }
	\label{fig:2b2mu_cross}
\end{figure}

To generate our signal and background events, we employ the following preselection cuts: 
\beq
\label{basic_cuts}
\Delta R_{jj, b \bar b, b \ell, j \ell } > 0.4, \quad 
\Delta R_{\ell \ell} > 0.2,  \quad 
p_{T}(j, b, \ell) > 10 ~{\rm GeV}, \quad 
|\eta_{j,b,\ell}| < 2.5 . 
\eeq 
After implementing these cuts, the dependence of the signal cross section on the coupling $g_{\mu\mu}$ is shown in Figure~\ref{fig:2b2mu_cross} for $M_{Z'} = 200, ~500 ~\rm{and}~ 1000 ~\rm{GeV}$. 
In our simulations, for a given $g_{\mu \mu}$ and $M_{Z'}$,  the value of $g_{bb}$ is fixed to the central value determined from Eq.~(\ref{eq:bestfit}).
The upper and lower ends of each signal cross-section curve are due to the finite allowed range of the couplings $g_{bb}$ and $g_{\mu\mu}$ as shown in Table \ref{benchmark1}.

We require the final state to be comprised of two oppositely charged muons 
and one or two $b$-tagged jets  with $p_T(b) > 20 $ GeV. 
We also impose an electron veto in the final state. 
The requirement of $b$-tagged jets helps to reduce the $\mu^+ \mu^- + jets$ background.
 The $p_T$ dependent $b$-tag efficiency ($\epsilon_b$) for the $b$-jets is  
 $\epsilon_b = 0.85 ~{\rm tanh}(0.0025~p_T)\Big(\cfrac{25.0}{1+0.063~ p_T}\Big)$. The misidentification efficiency functions for the $c$-jets ($\epsilon_c$) and that of the other light quark and gluon jets ($\epsilon_j$)  have the form,
 $\epsilon_c = 0.25~{\rm tanh}(0.018~ p_T)\Big(\frac{1}{1+ 0.0013 p_T}\Big)$ and $\epsilon_j = 0.01 + 0.000038~ p_T$ respectively \cite{Chatrchyan:2012jua}.
 
\begin{figure}[h]
	\centering
		\includegraphics[width=85mm,height=60mm]{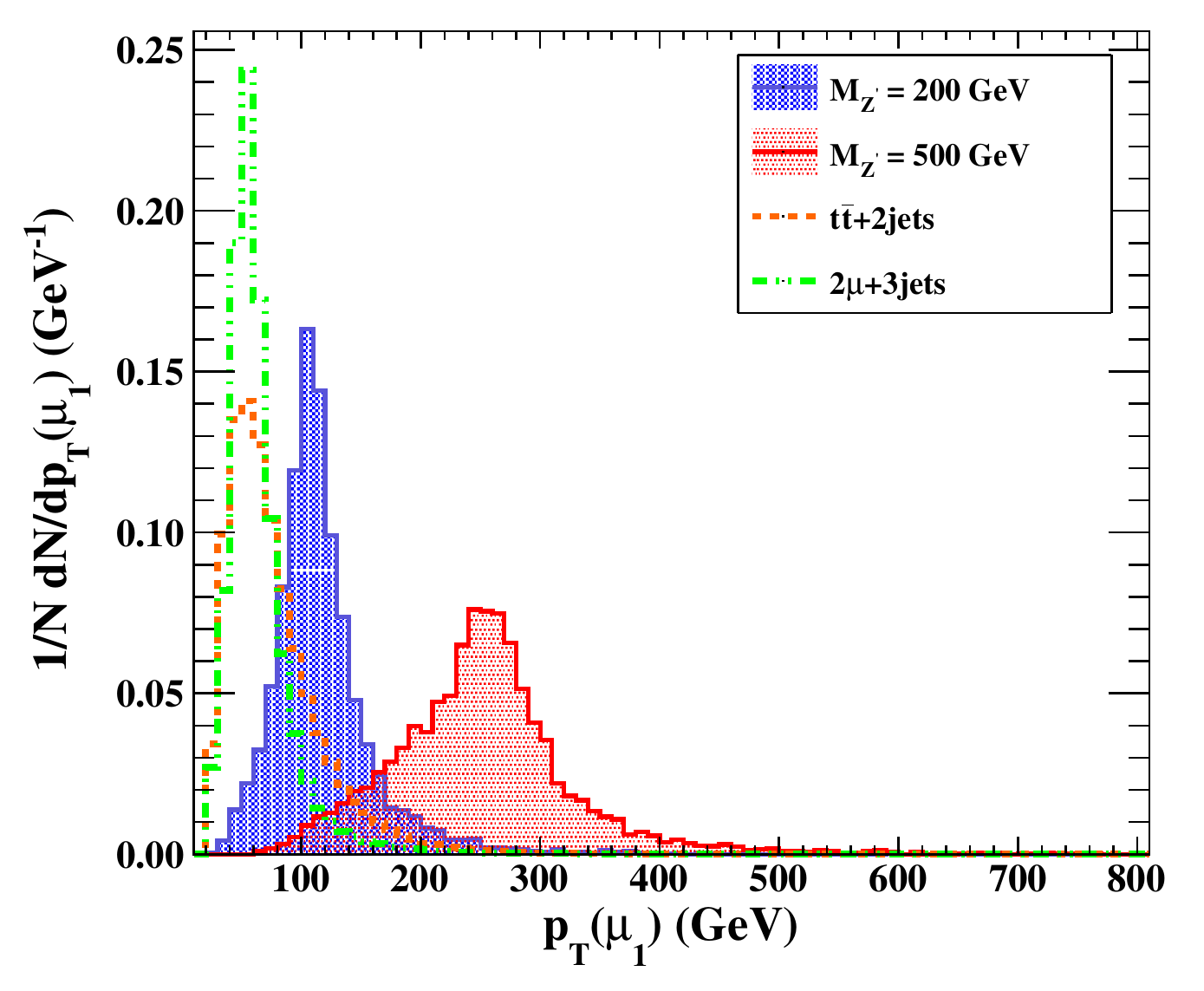} 
		\includegraphics[width=85mm,height=60mm]{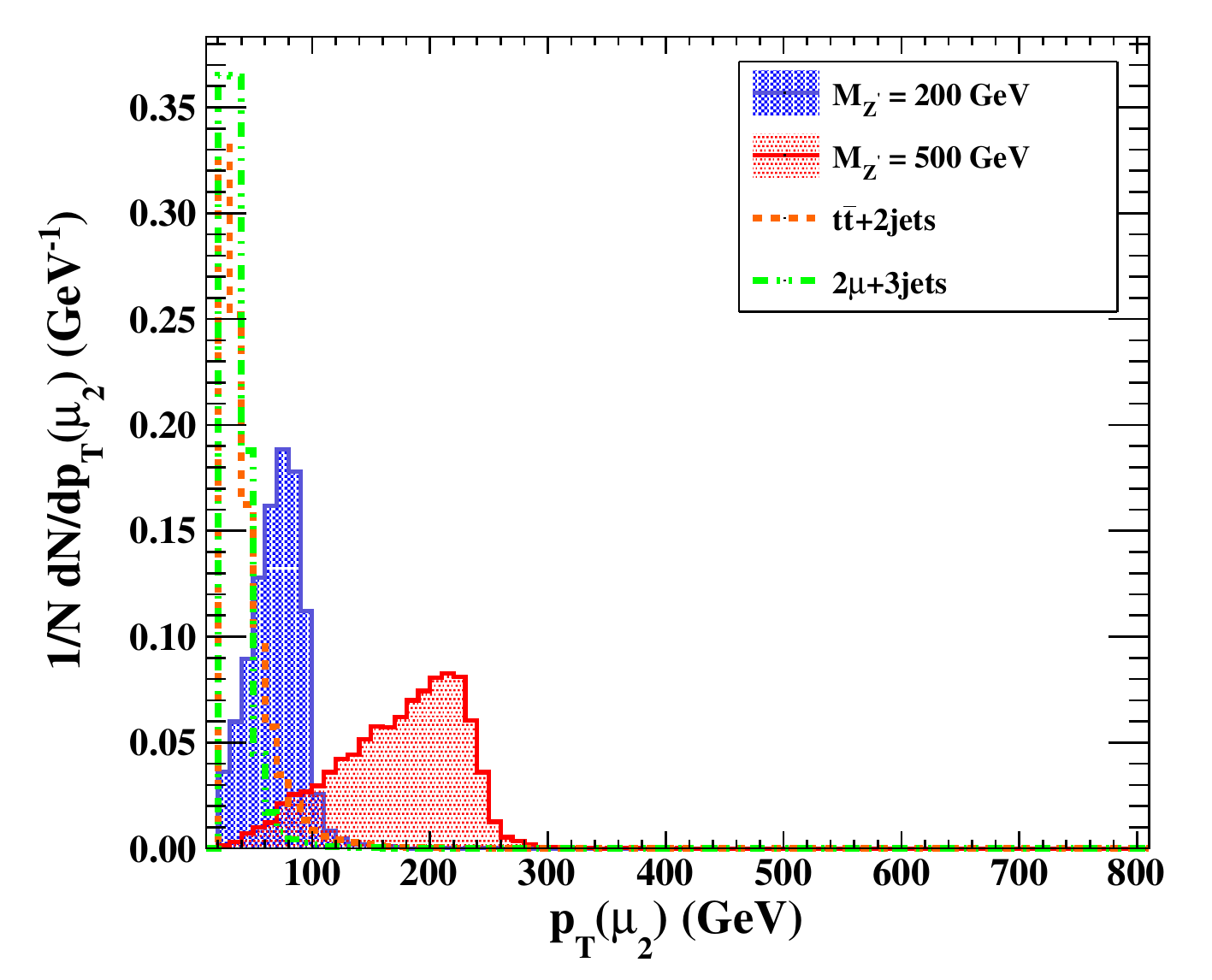}
	\caption{Normalized transverse momentum $(p_T)$ distributions of the leading (left) and sub-leading
	(right) muons for the signal ($M_{Z'} = 200~ \rm{and}~ 500$ GeV) and relevant SM backgrounds. The  values of $g_{\mu\mu}$ and $g_{bb}$ are 0.20(0.48) and 
	$4.2\times10^{-3}(1.10\times10^{-2})$ for $M_{Z'} = 200(500)$ GeV, respectively.}
	\label{fig:PT_mu} 
\end{figure}

To further optimize the signal selection cuts, we study the distributions of selected kinematic variables.
First, we study the transverse momentum distributions of the two muons. In the signal events these two muons originate from the decay of a heavy $Z^\prime $, while for the standard model background, they originate from the Drell-Yan process, from the decay of $t(\bar t)$ in top pair production process. For the signal, we show the distributions for two representative mass points $M_{Z^\prime} = 200 $ GeV and 500 GeV. Since the muons in the signal come from the decay of a heavy $Z'$, thus they are expected to have high transverse momentum. In comparison, the $p_T$ spectrum of muons for the SM background processes are expected to peak at relatively lower values.
In Figure \ref{fig:PT_mu},  the $p_T$ distributions of the  leading ($\mu_1$) and sub-leading ($\mu_2$) muons are contrasted between the signal and the background.   
We find that cutting on $p_{T}(\mu_1) > 90$ GeV, and $p_{T}(\mu_2) > 50$ GeV allows us to efficiently discriminate the signal over the SM background. 
 
%

%
%

%

%
%

\begin{figure}[h!]
	\centering
	\includegraphics[width=0.45\textwidth]{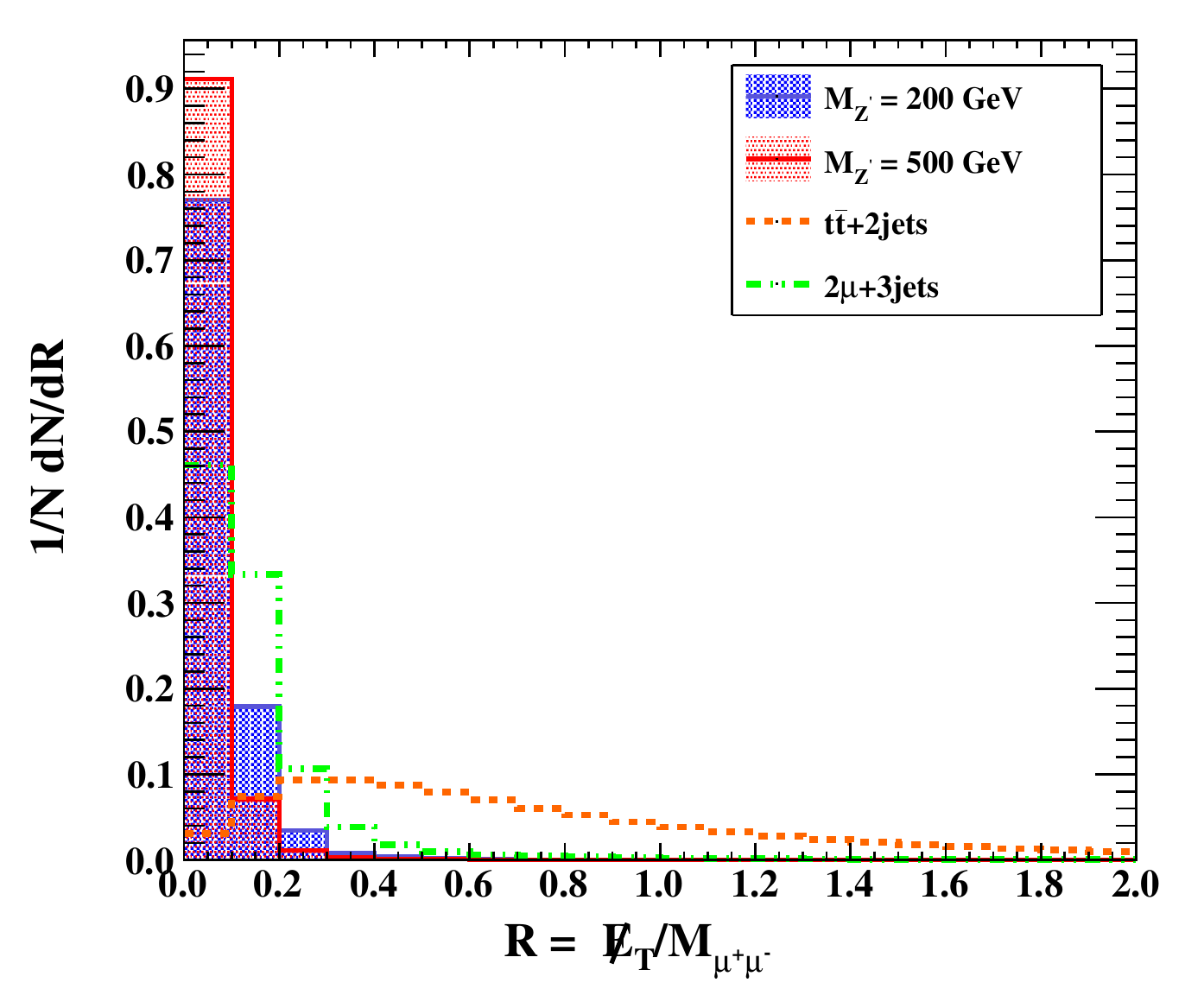}
	\caption{Normalized $R =  \cfrac{\cancel E_{T}}{M_{\mu^+ \mu^-}}$ distribution for signal and backgrounds.}
	\label{fig:ratio}
\end{figure}

\begin{figure}[h!]
	\centering
		\includegraphics[width=0.45\textwidth]{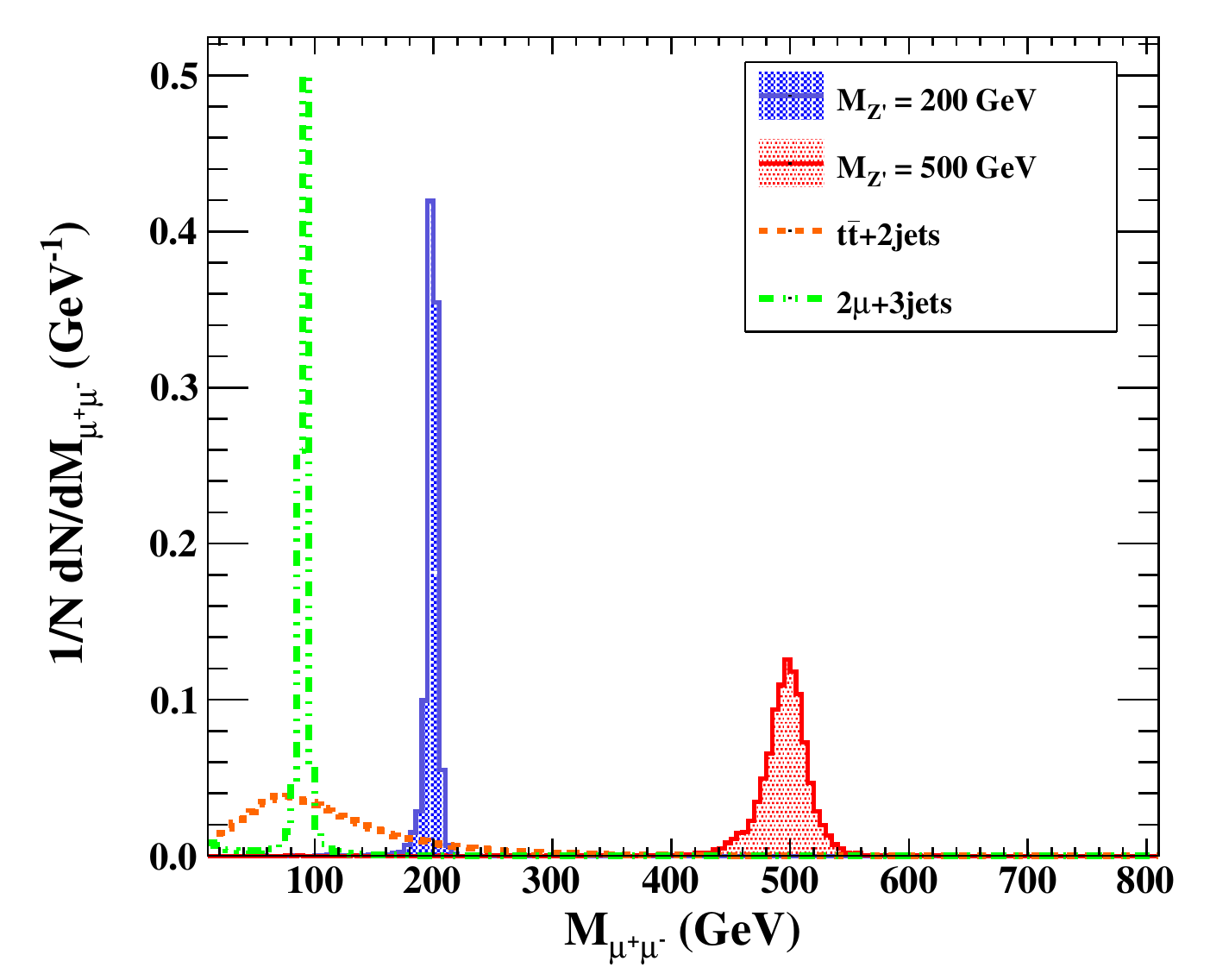} 
	
	\caption{ Normalized invariant mass distributions of the muon-pair for signal and backgrounds.  }
	\label{fig:IM_mumu}
\end{figure}

We now construct the kinematic variable $R$ defined as a ratio of the missing transverse 
energy $(\cancel E_{T})$ to the invariant mass of the muon-pair ($M_{\mu^+ \mu^-}$):
\begin{align}
R = \cfrac{\cancel E_{T}}{M_{\mu^+ \mu^-}}
\end{align}
For the signal, $\cancel E_{T}$ can come only from  $p_T$ mismeasurement of muons and $b$-jets, 
whereas  for the $t \bar t + jets$ background,  $\cancel E_{T} $ comes from the neutrinos in the leptonic decay of $W^\pm $. 
In Figure \ref{fig:ratio} we show the normalized distribution of $R$. For this reason, for the signal, $R$ peaks at a lower value while for  the $t \bar t$ background the distribution tends to peak at a higher value of $R$. 
We find that the cut $R< 0.2 $ allows one to significantly reduce the $t \bar t + jets$ background.
Finally we require the invariant mass of the muon pair to be in the window around the $Z'$ peak as dictated by,
\begin{align}
 |M_{\mu^+\mu^-}-M_{Z'}| < 6 \Gamma_{Z'} \nonumber
\end{align}
where  $\Gamma_{Z'}$ is the width of the $Z'$ resonance. This  cut is instrumental in further reducing the 
 $ \mu^+ \mu^- +  jets$ background as for these process the invariant mass of the muon pair
peaks around the $Z$ boson mass. 
The invariant mass distributions are depicted in Figure~\ref{fig:IM_mumu}.
%
%

%

Table~\ref{tab:bmu} summarizes the cuts discussed above and quantifies the effect of each cut on the signal and dominant backgrounds. 
Using these results, the signal significance can be determined from  the formula~\cite{Cowan:2010js}
\begin{align}
 \mathcal{S} = \sqrt{2\left[(S+B)\textrm{ln}\left(1+\frac{S}{B}\right)-S\right]}
 \label{signi_formula}
 \end{align}
 where $S(B)$ are the number of signal (background) events after all the cuts.  For calculating the
 significance, the signal and the backgrounds have been multiplied by respective k-factors to account for the next-to-leading-order (NLO) corrections. 
 For the signal we use the k-factor of 1.38 \cite{Campbell:2005zv}, while for $t \bar t + jets$ and $\mu^+ \mu^- + jets$ backgrounds we use the k-factors of 0.98 \cite{Bern:2008ef} and 1.15 \cite{Catani:2009sm},  respectively. 
 
 \begin{table}[h]
 \centering
 \footnotesize
\resizebox{17cm}{!}{
\begin{tabular}{|c|c|c|c|c|c|}
\hline
& \multicolumn{5}{|c|}{Cross-section after cut (fb)}
\\ \hline \hline
Process 
& Preselection &  $p_{T}(\mu_{1,2}) > 90(50)$~GeV  & $R < 0.2$~&~~ $|M_{\mu^+\mu^-}-200~{\rm GeV}| < 6 \Gamma_{Z'}$ ~~&~~ $|M_{\mu^+\mu^-}-500~{\rm GeV}| < 6 \Gamma_{Z'}$
\\ \hline 
$t\bar{t}+2jets$           
& 1861.04  & 264.38   & 78.00 & 3.80 & 1.42  
\\ \hline

$\mu^+ \mu^- + 3jets$
& 9438.13  & 317.08  & 168.38  & 2.47 & 0.31
\\ \hline
Total Background 
& 11299.17   & 581.46 &  246.38 & 6.27 &  1.73
\\ \hline \hline
 Signal: $M_{Z'}=200 ~\rm GeV$ 
& 1.76  & 1.22  & 1.15 & 0.92 & --  
\\ \hline 
 Signal:  $M_{Z'}=500 ~\rm GeV$
& 0.55 & 0.54 & 0.53 &-- & 0.39  
\\ \hline \hline
	\end{tabular}}
	\caption{ The  signal and background  cross sections for the $\mu^+ \mu^- + 1b(2b) $ process after each cut for $\sqrt{s} = 14$ TeV. The  values of $g_{\mu\mu}$ and $g_{bb}$ are $0.20(0.48)$ and $4.2\times10^{-3}(1.10\times10^{-2})$ for $M_{Z'} = 200(500)$ GeV, respectively.
	}
 \label{tab:bmu}
\end{table}
 
\begin{figure}[ht!]
\begin{subfigure}{.45\textwidth}
  \centering
  \includegraphics[width=0.9\linewidth]{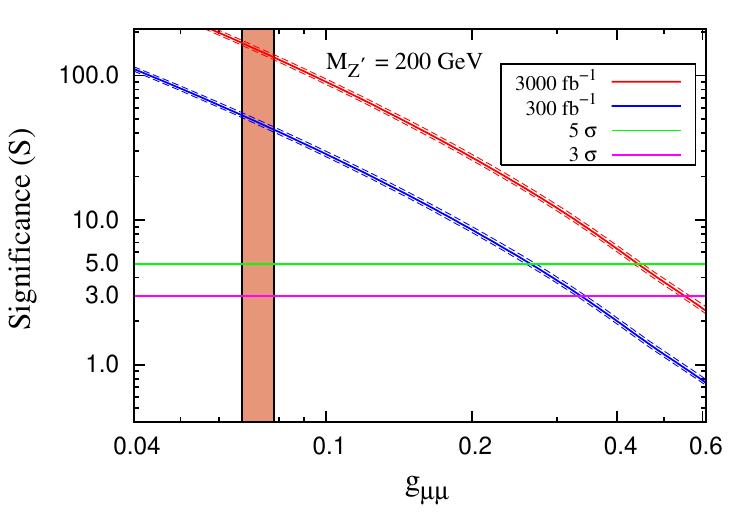}
  \caption{} \label{fig:sigma_200}
\end{subfigure}
\begin{subfigure}{.45\textwidth}
\centering
  \includegraphics[width=0.9\linewidth]{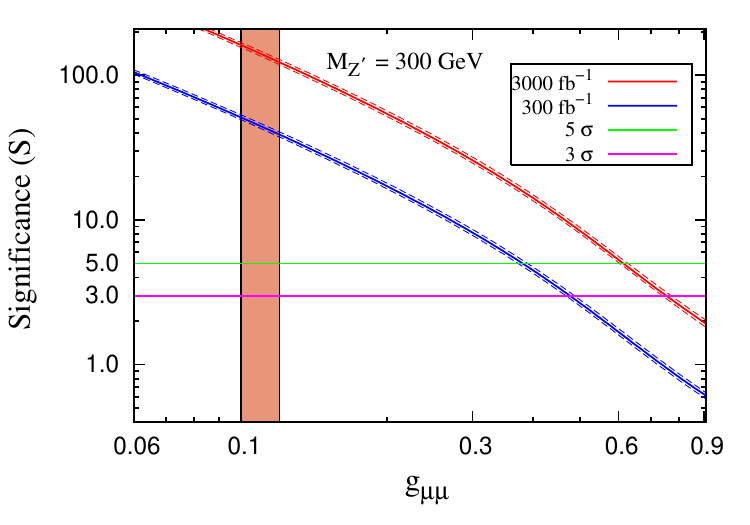}
  \caption{} \label{fig:sigma_300}
\end{subfigure}
\centering
\begin{subfigure}{.45\textwidth}
\centering
  \includegraphics[width=0.9\linewidth]{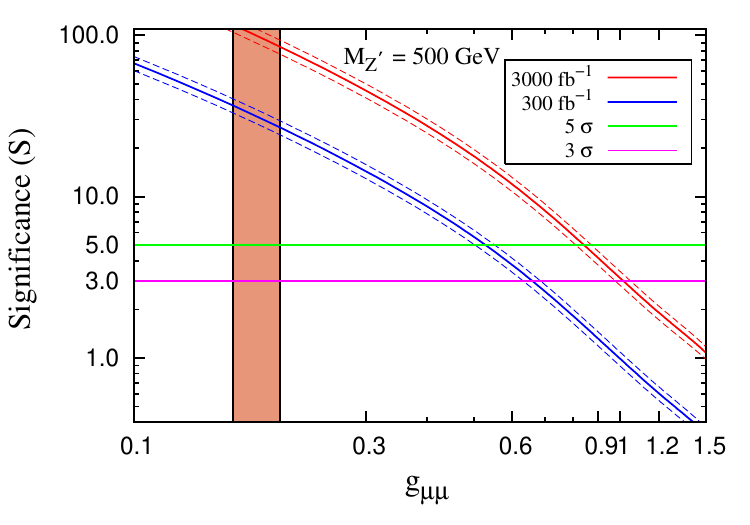}
  \caption{} \label{fig:sigma_500}
\end{subfigure}
\begin{subfigure}{.45\textwidth}
\centering
  \includegraphics[width=0.9\linewidth]{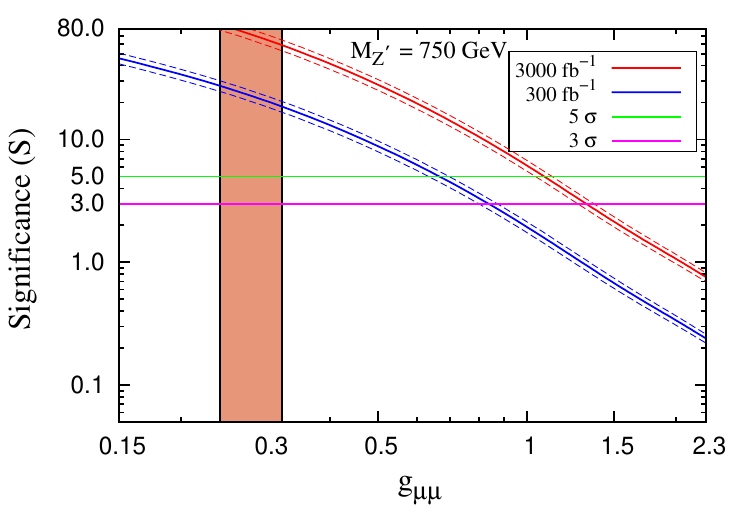}
  \caption{} \label{fig:sigma_750}
\end{subfigure}
\centering
   \begin{subfigure}{.45\textwidth}
  \centering
  \includegraphics[width=0.9\linewidth]{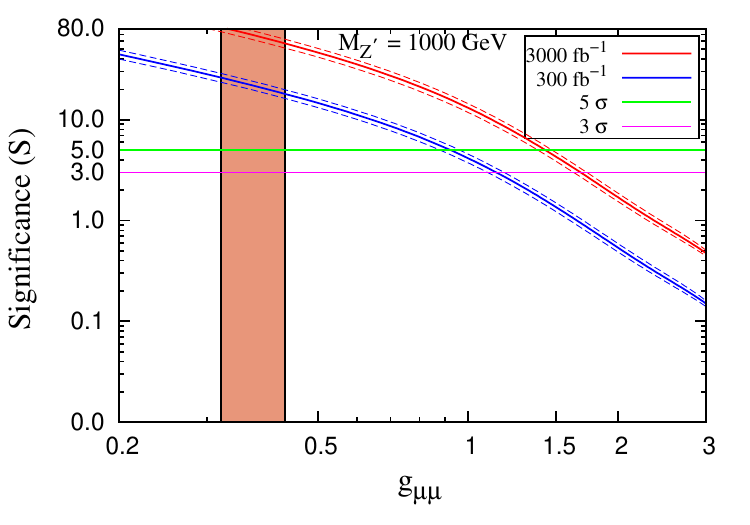}
  \caption{} \label{fig:sigma_1000}
\end{subfigure}%
 \caption{Significance vs. $g_{\mu\mu}$  for $M_{Z'} = 200 (\ref{fig:sigma_200})
 , ~300 (\ref{fig:sigma_300}) , ~500 (\ref{fig:sigma_500}), ~750 (\ref{fig:sigma_750}) 
 ~\rm{and} ~1000(\ref{fig:sigma_1000}) $ GeV for {$\mu^+ \mu^- + 1b(2b) $} channel at $\sqrt{s} = 14$ TeV. 
 The dashed lines represent the error band in Significance curves after including systematics $\sim 10 \%$
 in the background estimates. The dark shaded region is the one allowed
 at 1$\sigma$ CL by combining the constraints from $B$-meson anomalies, neutrino trident and $B \to D^{*} \ell \nu$.}
\label{fig:significance} 
 \end{figure}

 Based on the results in \tref{bmu}, we can calculate the signal significance for two particular benchmark points,  assuming the integrated luminosity of $300(3000)~{\rm fb}^{-1}$:
 \bea 
M_{Z'}= 200~{\rm GeV}, \quad g_{\mu \mu} = 0.20,\quad g_{bb} = 4.2\times10^{-3}: & \quad & S = 8.35 ~(26.4), 
\nnl 
M_{Z'}= 500~{\rm GeV}, \quad g_{\mu \mu} = 0.48, \quad g_{bb} = 1.1\times 10^{-2}: & \quad & S = 6.8 ~(21.5). 
 \eea 
 These benchmarks highlight the good prospect of observing the $Z'$ in this final state in the coming LHC runs. 
A broader set of results is shown in Figure~\ref{fig:significance}, 
where the signal significance for several representative values of $M_{Z'}$ is plotted as a function of the coupling $g_{\mu\mu}$.
One can see that the discovery potential in this final state is more more pronounced for lower $g_{\mu\mu}$ (which corresponds to higher $g_{bb}$).
As expected, the discovery potential quickly diminishes with the increasing $M_{Z'}$.
Nevertheless, a $5\sigma$ discovery is possible in this channel for $M_{Z'} \lesssim 500$~GeV with 300 fb$^{-1}$ luminosity at $\sqrt{s}=14$~TeV LHC, assuming the values of $g_{\mu \mu}$ and $g_{bb}$ preferred by the $b \to s \mu \mu$ anomalies and allowed by low-energy constraints. 
In the same conditions, a $3\sigma$ discovery is possible for $M_{Z'} \lesssim 1$~TeV.  


%


\subsection{$p p \to Z' \mu^{\pm} + \cancel{E}_T \to 3 \mu + \cancel{E}_T $ channel}
\label{sec:3muon}

We move to discussing another possible signature of the Z' particle: tri-muon plus missing energy final state. 
This final state in arises when the $Z'$ is radiated from 
$\mu^\pm$ or $\nu_{\mu}(\bar{\nu_{\mu}})$ leg in $pp \to W^{\pm *} \to Z' \mu^\pm  \nu_\mu ({\bar \nu_\mu}) $, followed by $Z^\prime \to \mu^+\mu^- $ decay.
As stated earlier, in this case both production and decay of the $Z'$ is controlled by its coupling $g_{\mu \mu}$ to the lepton sector. Thus this channel is best suited for probing the parameter space region with relatively higher values of $g_{\mu\mu}$.

\begin{figure}[ht!]
	\centering
		\includegraphics[width=0.7\textwidth]{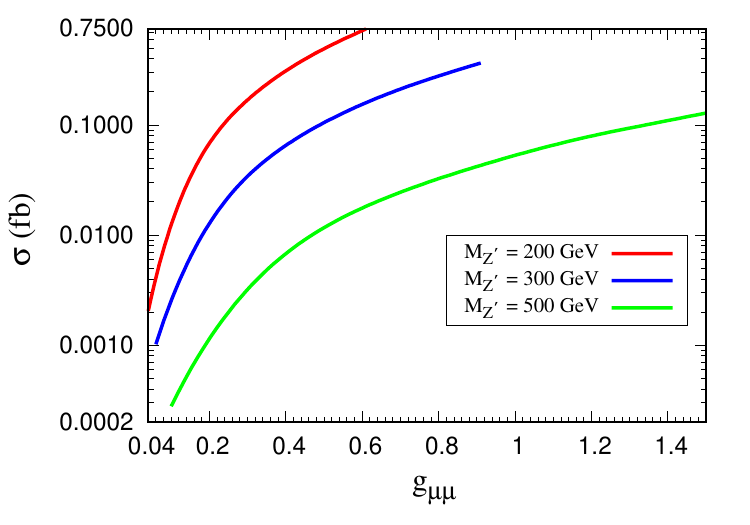} 
\caption{ 	\label{fig:3mumiss_cross}
Signal cross-section at $\sqrt{s} = 14$ TeV for 3$\mu + \cancel{E}_T$ channel as a function of $g_{\mu\mu}$ for $M_{Z'} = 200, ~ 300 ~\rm{and}~ 500 $  GeV.  }

\end{figure}

Similarly to the  $\mu^+ \mu^- + 1b(2b)$ analysis in the previous subsection,  we generate signal events in MadGraph with  the following preselection cuts: 
\beq
\label{basic_cuts_3mu_missET}
\Delta R_{jj, b \bar b, b \ell, j \ell } > 0.4, \quad  R_{\ell \ell} > 0.2, \quad 
p_{T}(j, b, \ell) > 10 ~{\rm GeV}, \quad 
|\eta_{j,b,\ell}| < 2.5. 
\eeq 
In Figure~\ref{fig:3mumiss_cross} we show the dependence of the leading order signal cross section of the coupling $g_{\mu\mu}$ after imposing the preselection cuts, for three representative values of $M_{Z'} = 200, ~ 300 ~\rm{and}~ 500 ~\rm{GeV}$. 

For the final state in question we can have the following SM processes that contribute to the background: $WZ$ + jets, $ZZ$ + jets, $WW$+jets, $t\bar t$,  $Z$ +jets.
Out of these,  $ W Z + {\rm jets}$ and $ Z Z + {\rm jets}$ are the irreducible backgrounds. 
$t\bar t$ can contribute to the background when each top quark decays leptonically: $t \to b \nu_\mu \mu$, and the third muon arises from the semileptonic decay of one of the b-quarks. 
Other sub-dominant contributions arise from $t \bar t V ~(V=W^{\pm}, ~Z)$ or $WWZ, WZZ$ channels \cite{Chatrchyan:2014aea}.

To optimize our signal versus background discrimination, we demand our final state to be comprised of exactly three muons with two muons of the same sign and the third muon of the opposite sign along with missing energy ($\cancel{E}_T$).  
We also impose a $b$-veto on the final state which helps us to reduce the $t\bar t$ background. For the two opposite sign dimuon pairs in the final state, we require their invariant masses, $M^{1}_{\rm OSD},~M^{2}_{\rm OSD}$ 
to satisfy
\begin{align}
  M^{1,2}_{\rm OSD} < 75~{\rm GeV}~  ~ {\rm or } ~ ~M^{1,2}_{\rm OSD} > 105~{\rm GeV}.
\end{align}
This helps to exclude the background contribution where the opposite sign muon pair(s) arise from $Z$ resonance. We also
impose $M^{1,2}_{\rm OSD} > 12$ GeV to suppress the Drell-Yan background~\cite{Chatrchyan:2014aea}.
With the above criteria, the dominantly surviving background contribution comes from $W Z$+jets\footnote{For validation, we have generated the $WZ$ background for $\surd s =$ 13 TeV using 
Madgraph@NLO~\cite{Alwall:2014hca} and compared with
the background event expectation as given in the CMS paper~\cite{Sirunyan:2017lae}. Within $1\sigma$ and $2\sigma$ uncertainty of our Monte Carlo simulation, we
are consistent with the CMS background numbers up to 15$\%$ and 11$\%$ respectively.}.

\begin{figure}[ht!]
\begin{subfigure}[b]{.45\textwidth}
  \includegraphics[width=0.9\linewidth]{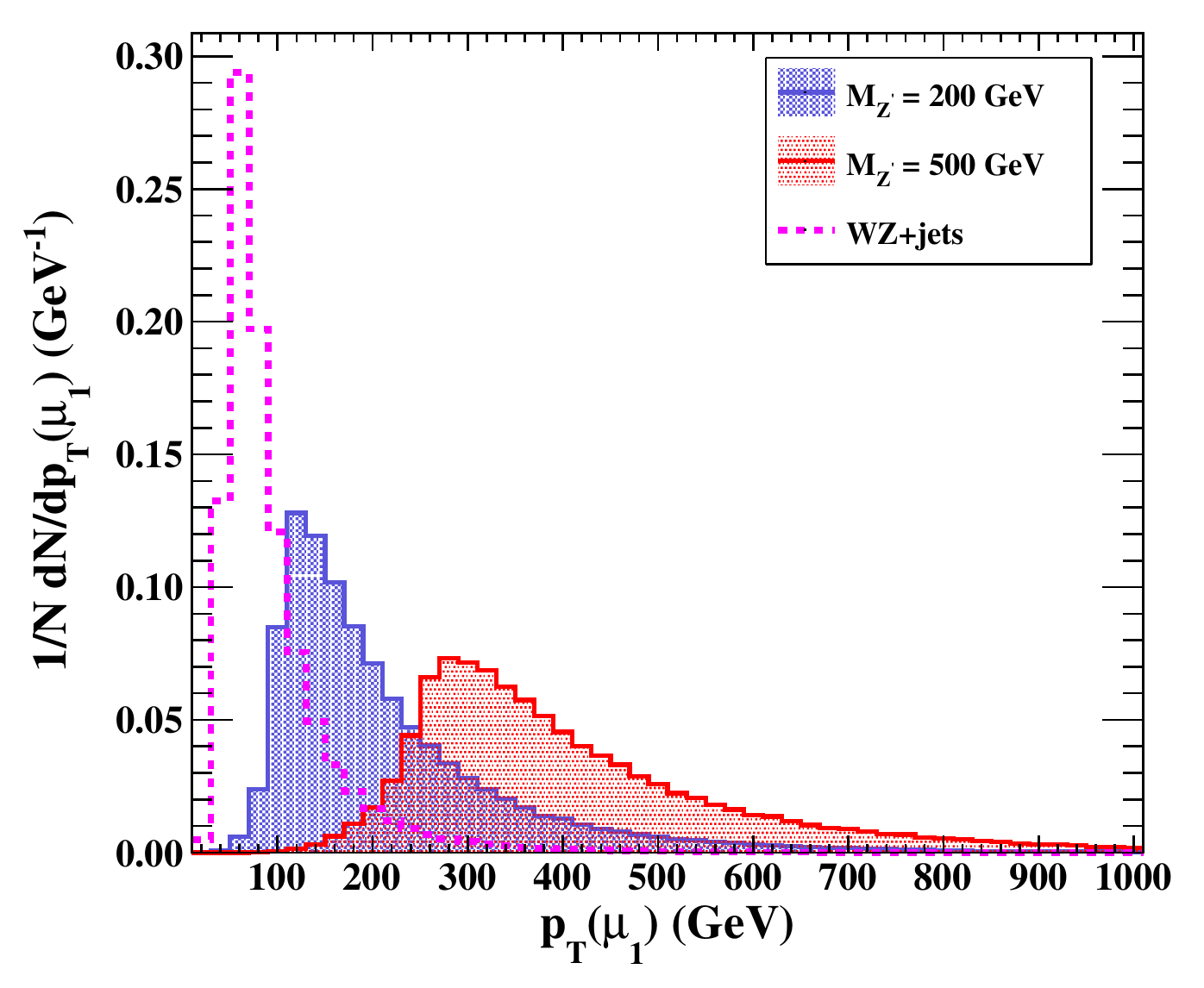}
  \caption{} \label{fig:3mu_misset_pt_mu1}
\end{subfigure}%
     \begin{subfigure}[b]{.45\textwidth}
  \centering
  \includegraphics[width=0.9\linewidth]{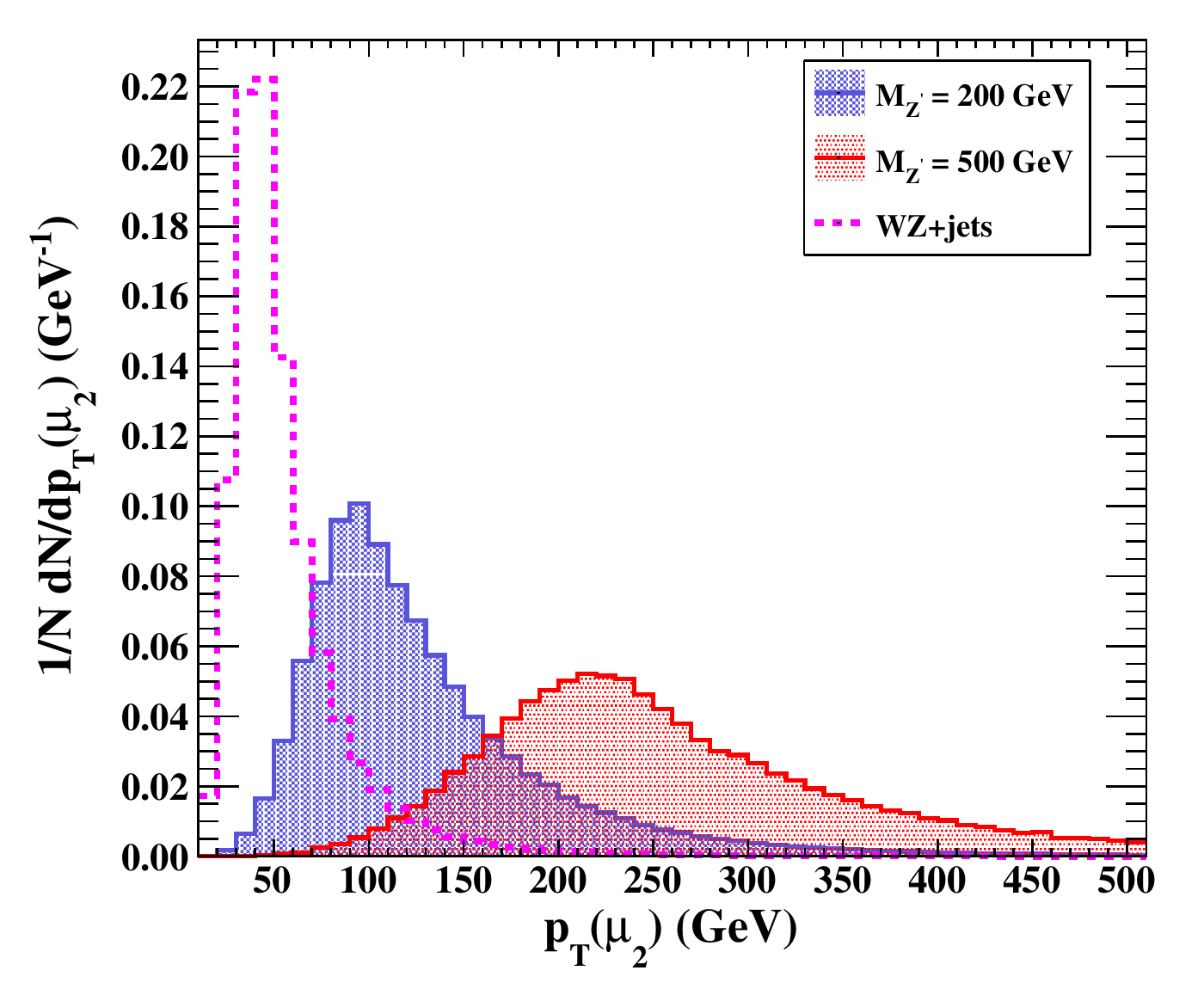}
  \caption{} \label{fig:3mu_misset_pt_mu2}
\end{subfigure}%

\centering
   \begin{subfigure}[b]{.45\textwidth}
  \includegraphics[width=0.9\linewidth]{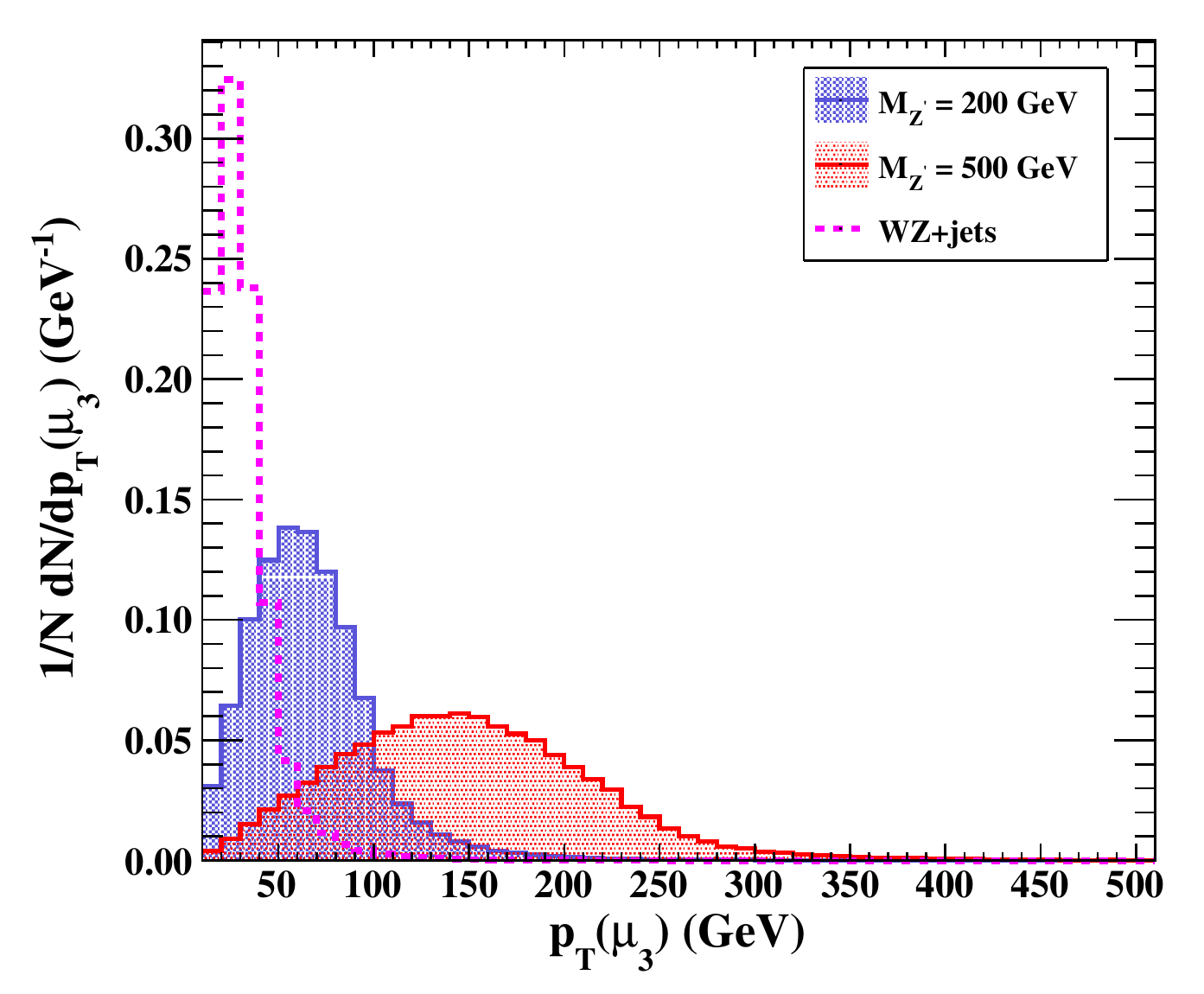}
  \caption{} \label{fig:3mu_misset_pt_mu3}
\end{subfigure}%
\caption{\label{fig:PT_mu_3mu} 
Normalized transverse momentum $(p_T)$ distributions of the leading (\ref{fig:3mu_misset_pt_mu1}),
sub-leading (\ref{fig:3mu_misset_pt_mu2}) and sub-sub-leading (\ref{fig:3mu_misset_pt_mu3}) muons
for the $3\mu + \cancel{E}_T$ final state. 
Signal distributions are for $M_{Z'} = 200$ GeV, $g_{\mu\mu} = 0.20$, $g_{bb} = 4.2\times10^{-3}$, 
and for $M_{Z'} = 500$ GeV,	 $g_{\mu\mu} = 0.48$, $g_{bb} = 1.10\times10^{-2}$.
We also show the analogous distributions for the $WZ$ background.
}
\end{figure}

In our analysis we assume that the $Z'$ mass is greater than the $Z$ and $W^{\pm}$ boson masses. 
Thus, the muons in the signal are expected to have higher $p_T$  than those coming from the decay of the $Z$ or $W^{\pm}$ bosons in the SM backgrounds. The comparative distributions of the transverse momenta of the the leading ($\mu_1$), sub-leading ($\mu_2$) and sub-sub-leading muons ($\mu_3$) in the final state for the signal and backgrounds are shown in Figure ~\ref{fig:PT_mu_3mu}. 
To enhance the signal over background ratio we impose the following cuts 
\begin{align}
  p_{T}(\mu_1) > 100~ {\rm GeV},~~ p_{T}(\mu_2) > 70 ~ {\rm GeV},~~ p_{T}(\mu_3) > 40~ {\rm GeV}. 
\end{align}

\begin{figure}[h]
	\centering
 	\begin{tabular}{cc}
		
		\includegraphics[width=0.4\linewidth]{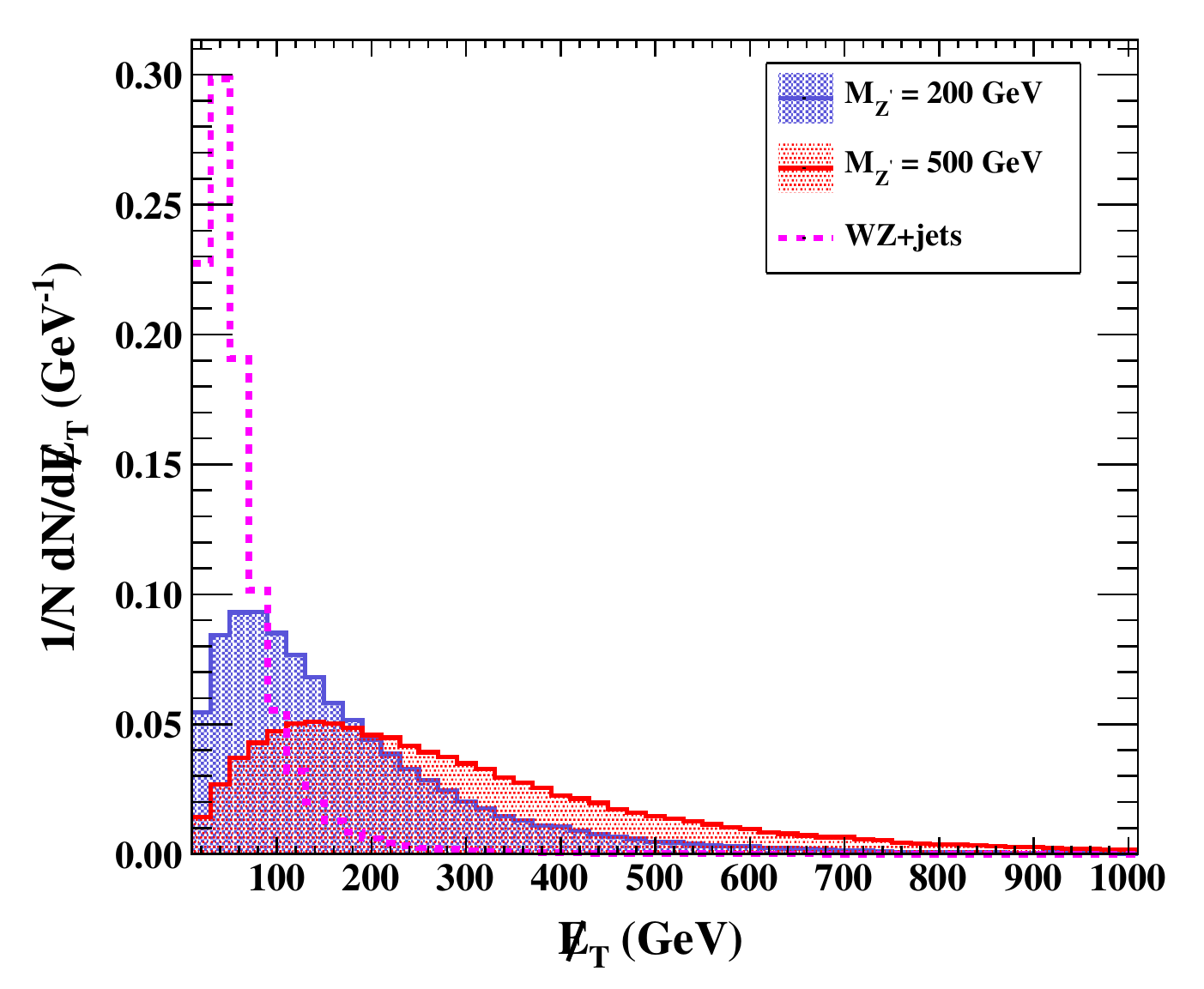}
 	\end{tabular}
	\caption{ 
	Normalized missing energy ($\cancel{E}_T$) distribution for the $3\mu + \cancel{E}_T$ final state. 
	We show the distribution for the signal for $M_{Z'} = 200$ GeV, $g_{\mu\mu} = 0.20$, $g_{bb} = 4.2\times10^{-3}$, and for  $M_{Z'} = 500$ GeV,  $g_{\mu\mu} = 0.48$, $g_{bb} = 1.1\times10^{-2}$.
We also show the analogous distributions for the $WZ$ background.	
	}
	\label{fig:missET_3mu_missET} 
\end{figure}

Finally, in Figure ~\ref{fig:missET_3mu_missET} we compare the $\cancel{E}_T$  distributions of signal and the $WZ$+jets background. 
The missing energy for the background comes from the leptonic decay of the $W^{\pm}$ boson in  $WZ  + {\rm jets}$ or from mismeasurement of leptons or jets in the Drell-Yan process. 
As a result, the distribution of $\cancel{E}_T$ peaks at around half of the $W^{\pm}$ mass for the background, 
whereas for the signal it is shifted towards higher values. 
In our analysis we impose the cut $\cancel{E}_T >$ 60 GeV which provides an optimal cut capturing the relatively long tail in the signal and avoiding the peak in the $WZ$+jets background.

We summarize the above discussed cut flow in Table~\ref{tab:3mumet} for our two representative benchmark points.
Given these results, we can calculate the signal significance for our 2 benchmark points, assuming the integrated luminosity of $300(3000)~{\rm fb}^{-1}$:
 \bea 
M_{Z'}= 200~{\rm GeV}, \quad g_{\mu \mu} = 0.20,\quad g_{bb} = 4.2\times10^{-3} : & \quad & S = 1.6 ~(5.0), 
\nnl 
M_{Z'}= 500~{\rm GeV}, \quad g_{\mu \mu} = 0.48, \quad g_{bb} = 1.1\times 10^{-2}: & \quad & S = 0.4 ~(1.3). 
 \eea 

\begin{table}[h]
 	\centering
 	\footnotesize
 	\resizebox{13cm}{!}{
 		\begin{tabular}{|c|c|c|c|c|}
 			\hline
 			\multicolumn{1}{|c|}{}& \multicolumn{4}{|c|}{Effective Cross-section after each cut (fb)}  \\ \hline
 		Process	& Preselection & $M^{1,2}_{\rm OSD}$ cut ~&~$ p_{T}(\mu)$ cut & $\mET$ cut ~~ 
 			\\ \hline \hline 
Background: $WZ$+2j & 107.02  & 3.31 & 0.31 & 0.15 \\ \hline \hline
Signal: $M_{Z}'=200 ~\rm GeV$ & $6.5\times 10^{-2}$  &  $5.9\times 10^{-2}$  			&  $4.4\times 10^{-2}$  &  $3.7\times 10^{-2}$ \\ \hline
Signal: $M_{Z}'=500 ~\rm GeV$ &  $1.01\times 10^{-2}$ & $9.8\times 10^{-3}$ &$9.56\times 10^{-3}$ 
 			& $9.0\times 10^{-3}$  \\ \hline \hline
 	\end{tabular} }
 	\caption{ Effective cross-section at $\sqrt{s} = 14$ TeV for both signal and background for $3 \mu + \mET $ channel
 	after each cut described in the next . 
The signal benchmarks correspond to the couplings $g_{\mu\mu} = 0.20$, $g_{bb} = 4.2\times10^{-3}$ for $M_{Z'} = 200$ GeV, and   $g_{\mu\mu} = 0.48$, $g_{bb} = 1.10\times10^{-2}$ for  $M_{Z'} = 500$ GeV. 	
 	}
  \label{tab:3mumet}
  \end{table}

\begin{figure}[ht!]
\begin{subfigure}{.45\textwidth}
  \centering
  \includegraphics[width=0.9\linewidth]{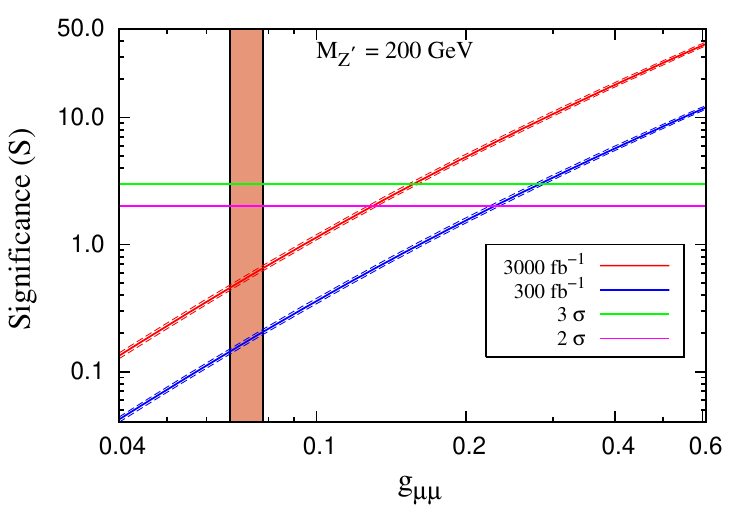}
  \caption{} \label{fig:sigma_200_3mumisset}
\end{subfigure}%
     \begin{subfigure}{.45\textwidth}
  \centering
  \includegraphics[width=0.9\linewidth]{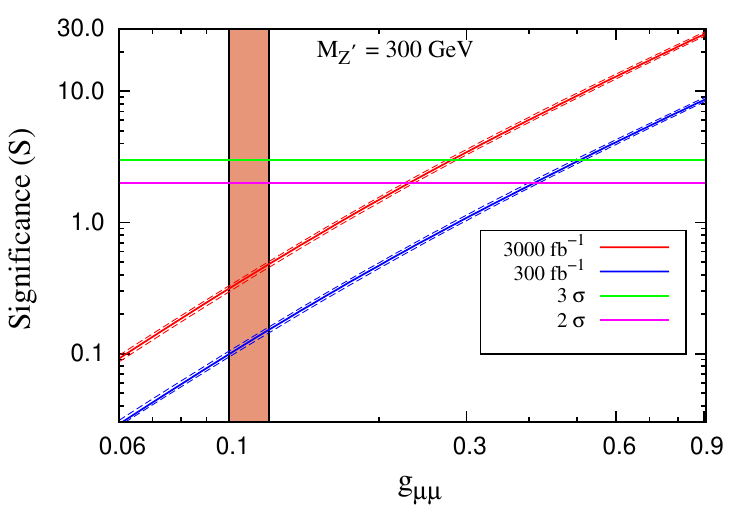}
  \caption{} \label{fig:sigma_300_3mumisset}
\end{subfigure}%

\centering
   \begin{subfigure}{.45\textwidth}
  \centering
  \includegraphics[width=0.9\linewidth]{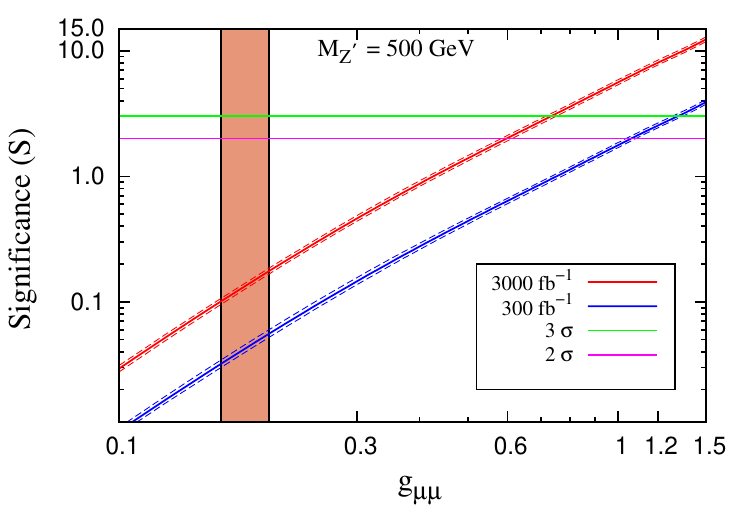}
  \caption{} \label{fig:sigma_500_3mumisset}
\end{subfigure}%

 \caption{Significance in the $3\mu + \cancel{E}_T$ channel as a function of  $g_{\mu\mu}$  for $M_{Z'} = 200 (\ref{fig:sigma_200_3mumisset})
 , ~300 (\ref{fig:sigma_300_3mumisset}) ~\rm{and} ~500 (\ref{fig:sigma_500_3mumisset})$ GeV 
 for $\sqrt{s} = 14$ TeV.  
 The dashed lines represent the error band for the  significance curves after including systematics $\sim 10 \%$  in the background estimates. 
 The dark shaded region is the one allowed
 at 1$\sigma$ CL by combining the constraints from $B$-meson anomalies, neutrino trident and $B \to D^{*} \ell \nu$.}
\label{fig:significance_3mumiss}
 \end{figure}

The projected significance for our analysis in the $3\mu + \cancel{E}_T$  channel  as a function of the coupling $g_{\mu\mu}$ is portrayed for $M_{Z'} = 200 (\ref{fig:sigma_200_3mumisset}), ~300 (\ref{fig:sigma_300_3mumisset}) ~\rm{and} ~500 (\ref{fig:sigma_500_3mumisset})$ GeV in Figure~\ref{fig:significance_3mumiss}.
For the significance calculation signal and background have been scaled by k-factors of 1.25 \cite{ATLAS:2016pbt} and 1.83 \cite{Grazzini:2016swo} respectively.
Note that in this case, and unlike in the previously discussed $\mu^+ \mu^- +1b(2b)$ channel, the significance increases with increasing $g_{\mu\mu}$. 
This demonstrates the complementarity of the two final states discussed in this paper. 

\section{Comparison with dimuon searches}
\label{sec:dimuon}
\setcounter{equation}{0}

Our $Z'$ model leads to additional LHC signatures besides those studied in Sections~\ref{sec:mumubb} and \ref{sec:3muon}.  
One is the 4 muon final state produced in the process $pp \rightarrow Z \rightarrow 4 \mu$ where the Z boson decays to a muon pair and an on-shell or virtual $Z'$ is radiated off a muon and subsequently decays into pair of muons.
This is however relevant only for fairly low Z' masses,   $5\lesssim M_{Z'} \lesssim 70$~GeV~\cite{Altmannshofer:2014cfa,Altmannshofer:2014pba,Altmannshofer:2016jzy}, which are  outside of  our direct interest in this paper.     
For a heavier $Z'$, the strongest constraints comes from dimuon resonance searches, $pp\rightarrow Z'\rightarrow \mu^- \mu^+$~\cite{Dalchenko:2017shg,Allanach:2018odd}. 
In our scenario, $Z'$ is dominantly produced through its couplings to bottom quarks.  Its  branching fraction into muons depends on $g_{\mu \mu}$, $g_{bb}$ and $M_{Z'}$, and it is typically significant in the interesting parameter space of the model. Other than to muons, $Z'$ may also decay into top and bottom quarks and into neutrinos, however these channels are less competitive.  In particular, we have verified that the constraints from dijet resonance searches at the LHC~\cite{Aaboud:2017yvp,CMS:2017xrr} are much weaker than those we obtain from the dimuon resonance searches. 

\begin{figure}[h!]
\begin{subfigure}[b]{.48\textwidth}
  \includegraphics[width=0.9\linewidth]{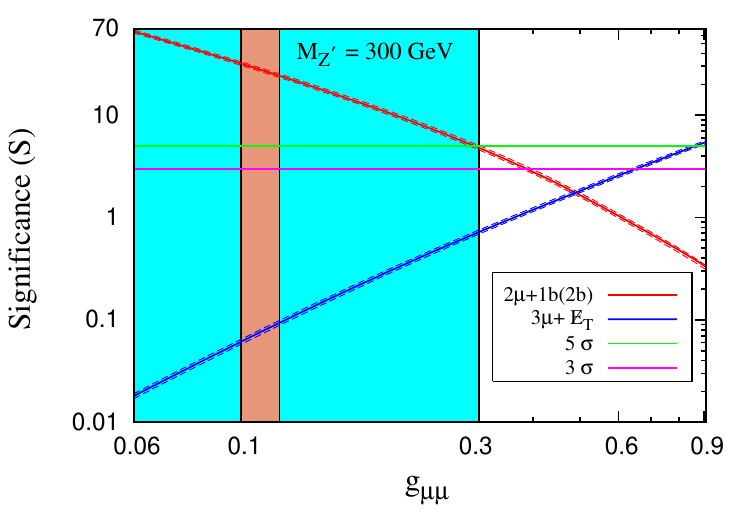}
  \caption{} \label{fig:13TeV_Z'bb_MZp_300}
\end{subfigure}%
     \begin{subfigure}[b]{.48\textwidth}
  \centering
  \includegraphics[width=0.9\linewidth]{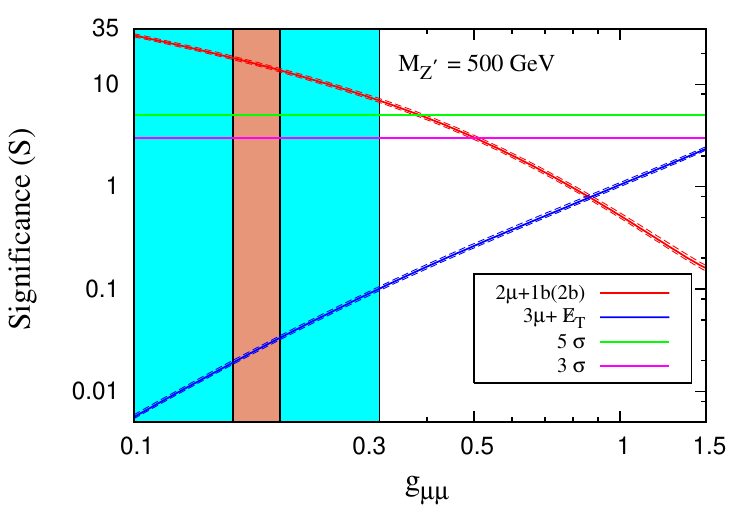}
  \caption{} \label{fig:13TeV_Z'bb_MZp_500}
\end{subfigure}%

\caption{
The parameter range of our model excluded at 95$\%$ CL by the ATLAS dimuon resonance search~\cite{ATLAS:2019vcr,Aad:2019fac} at $\sqrt{s} = 13$ TeV with 139~ fb${}^{-1}$ (light blue band) for  $M_{Z'} =300$ GeV (\ref{fig:13TeV_Z'bb_MZp_300}) and $M_{Z'} =500$ GeV (\ref{fig:13TeV_Z'bb_MZp_500}). 
This is compared with the signal significance expected in the $ 2 \mu+ 1b(2b)$ (red) and $3 \mu + \mET$ (blue) channels for the same collision energy and luminosity. 
The brown region is preferred at 1$\sigma$ CL by combining the constraints from $b\to s \ell \ell$ anomalies, the neutrino trident production, and $B \to D^{*} \ell \nu$ processes.}
\label{fig:13TeV_Z'bb} 
\end{figure}

Figure~\ref{fig:13TeV_Z'bb} illustrates constraints on the parameter space of the model from dimuon resonance searches. 
The blue band shows the range of $g_{\mu \mu}$ excluded at 95$\% $ CL by the ATLAS analysis at 13 TeV with 139 fb${}^{-1}$ of data~\cite{ATLAS:2019vcr,Aad:2019fac}. 
We show the exclusion region for $M_{Z'} = 300$~GeV and $500$~GeV, assuming the coupling $g_{bb}$ is determined by the central value of the best fit to the  $b\to s \ell \ell$ anomalies in~\eref{bestfit}.  
We can see that the regions with smaller $g_{\mu\mu}$ (hence larger $g_{bb}$) are disfavored; in particular the region preferred by the global fit to low-energy data is excluded by the LHC. 
Nevertheless, an important chunk of the parameter space remains allowed at $2\sigma$ by all existing LHC and low-energy analyses.
Those region will be probed in the future LHC runs.  

Furthermore, from Figure~\ref{fig:13TeV_Z'bb} we learn that the dimuon and $2\mu+b$ searches probe similar regions of the parameter space, and they exhibit a similar sensitivity.
This is not an accident, as the two signals are closely related, and there is an overlap between the dimuon resonance and the $2\mu+b$ signal regions.  
We note however that dimuon resonances are predicted by multiple new physics scenarios. 
Conversely,  observing a signal in the $2\mu+b$ channel would be a spectacular confirmation that the newly found resonance could explain the  $b\to s \ell \ell$ anomalies. 

On the other hand, in Figure~\ref{fig:13TeV_Z'bb} we observe that the $3 \mu + \mET$ process probes a complementary region of the parameter space compared to the $2\mu+b$ channel or generic dimuon resonance searches.    
Combining information from all of these channels will allow one to completely exclude  the parameter space  of our model with $M_{Z'} \lesssim 500$~GeV. 
Heavier $Z'$ resonances may escape discovery at the LHC in the parameter space preferred by the $b\to s \ell \ell$ anomalies.


 \section{Summary and Conclusions}\label{Summary}
 
 In this work we have analyzed the LHC discovery prospects of a new massive spin-1 particle ($Z'$) in a model explaining the $b \to s \ell^+ \ell^-$ anomalies. 
We focused on the model proposed in Ref.~\cite{Falkowski:2018dsl} where  tree-level exchange of the $Z'$ contributes to  $b \to s \mu^+ \mu^-$ processes, and can explain in  particular the apparent violation of lepton flavor universality encoded in the $R_{K^{(*)}}$ observables. 
In this model the $Z'$ has sizable couplings to left-handed bottom quarks and muons, as well as their $SU(2)_W$ partners. 
Therefore it can be produced on its own via $b \bar b$ fusion and decay into a muon pair, showing up at the LHC as a dimuon resonance. 
In addition, the $Z'$ can be produced in association with another SM particle when it is radiated off a bottom, top, muon, and neutrino legs. 
While the dimuon resonance signature has been previously studied in this context, the associated  production is less explored. 
In this paper we identified two promising signatures of the associated $Z'$ production:   $p p \to Z'+ 1(2)b$ with $Z'$ radiated of a bottom quark, and $p p \to Z' \mu^{\pm} + \cancel{E}_T $ with $Z'$ radiated of a muon or a neutrino.  
In both cases we focused on $Z'$ decays to $\mu^+ \mu^-$. 

The interesting parameter space of our model  can be succinctly characterized by two variables: the $Z'$ mass $M_{Z'}$, and its  coupling to muons $g_{\mu \mu}$. 
The coupling to b-quarks $g_{bb}$ is approximately fixed by the previous two via \eref{bestfit} as a result of fitting the $b \to s \ell^+ \ell^-$ anomalies. 
From \eref{bestfit} $g_{bb}$ and $g_{\mu \mu}$ are anti-correlated. 
We find that the $p p \to Z'+b$ channel is sensitive to lower values of $g_{\mu\mu})$, as the $Z'$ production cross section is proportional $g_{bb}^2$. 
This feature is the same as for $Z'$ produced alone, and we find that these two production mechanisms offer a comparable sensitivity to the parameter space of the model.  
Conversely, $p p \to Z' \mu^{\pm} + \cancel{E}_T $ is sensitive to larger $g_{\mu\mu})$ as the $Z'$ production cross section is proportional $g_{\mu\mu}^2$. 
Taken together, the two associated production channels offer a good and complementary sensitivity to a wide range of the parameter space explaining the $b \to s \ell^+ \ell^-$ anomalies for $200 \lesssim M_{Z'} \lesssim 500$~GeV.


 \section{Acknowledgement}
 NG would like to acknowledge the Council of Scientific and Industrial Research (CSIR), Government
of India for financial support. NG and SD would like to thank Professor Satyaki Bhattacharya, Abhaya kumar Swain and Disha Bhatia
for useful discussions.
This work was supported in part by the CNRS LIA (Laboratoire International Associé) THEP (Theoretical High Energy Physics) and the INFRE-HEPNET (IndoFrench Network on High Energy Physics) of CEFIPRA/IFCPAR (Indo-French Centre for the Promotion of Advanced Research).
A.F. is partially supported by the European Union’s Horizon 2020 research and innovation programme under the Marie Sklodowska-Curie grant agreements No 690575 and No 674896.
DKG wishes to acknowledge the hospitalities 
of the LPT Orsay where this project was initiated and the Theoretical Physics Department, CERN, Switzerland, where part of this project was completed.
 
\bibliographystyle{JHEP}  
\bibliography{ref1}

\providecommand{\href}[2]{#2}\begingroup\raggedright\begin{thebibliography}{10}

\bibitem{Aaij:2019wad}
{\bf LHCb} Collaboration, R.~Aaij et~al., {\it {Search for lepton-universality
  violation in $B^+\to K^+\ell^+\ell^-$ decays}},  {\em Phys. Rev. Lett.} {\bf
  122} (2019), no.~19 191801, [\href{http://arxiv.org/abs/1903.09252}{{\tt
  arXiv:1903.09252}}].

\bibitem{Aaij:2017vbb}
{\bf LHCb} Collaboration, R.~Aaij et~al., {\it {Test of lepton universality
  with $B^{0} \rightarrow K^{*0}\ell^{+}\ell^{-}$ decays}},  {\em JHEP} {\bf
  08} (2017) 055, [\href{http://arxiv.org/abs/1705.05802}{{\tt
  arXiv:1705.05802}}].

\bibitem{Alguero:2019ptt}
M.~Algueró, B.~Capdevila, A.~Crivellin, S.~Descotes-Genon, P.~Masjuan,
  J.~Matias, and J.~Virto, {\it {Addendum: "Patterns of New Physics in $b\to s
  \ell^+\ell^-$ transitions in the light of recent data" and "Are we
  overlooking Lepton Flavour Universal New Physics in $b \to s \ell\ell\,$?"}},
   \href{http://arxiv.org/abs/1903.09578}{{\tt arXiv:1903.09578}}.

\bibitem{Ciuchini:2019usw}
M.~Ciuchini, A.~M. Coutinho, M.~Fedele, E.~Franco, A.~Paul, L.~Silvestrini, and
  M.~Valli, {\it {New Physics in $b \to s \ell^+ \ell^-$ confronts new data on
  Lepton Universality}},  \href{http://arxiv.org/abs/1903.09632}{{\tt
  arXiv:1903.09632}}.

\bibitem{Aebischer:2019mlg}
J.~Aebischer, W.~Altmannshofer, D.~Guadagnoli, M.~Reboud, P.~Stangl, and D.~M.
  Straub, {\it {$B$-decay discrepancies after Moriond 2019}},
  \href{http://arxiv.org/abs/1903.10434}{{\tt arXiv:1903.10434}}.

\bibitem{Kowalska:2019ley}
K.~Kowalska, D.~Kumar, and E.~M. Sessolo, {\it {Implications for New Physics in
  $b\to s \mu\mu$ transitions after recent measurements by Belle and LHCb}},
  \href{http://arxiv.org/abs/1903.10932}{{\tt arXiv:1903.10932}}.

\bibitem{Arbey:2019duh}
A.~Arbey, T.~Hurth, F.~Mahmoudi, D.~Martinez~Santos, and S.~Neshatpour, {\it
  {Update on the b->s anomalies}},  \href{http://arxiv.org/abs/1904.08399}{{\tt
  arXiv:1904.08399}}.

\bibitem{Gauld:2013qba}
R.~Gauld, F.~Goertz, and U.~Haisch, {\it {On minimal $Z'$ explanations of the
  $B\to K^*\mu^+\mu^-$ anomaly}},  {\em Phys. Rev.} {\bf D89} (2014) 015005,
  [\href{http://arxiv.org/abs/1308.1959}{{\tt arXiv:1308.1959}}].

\bibitem{Buras:2013qja}
A.~J. Buras and J.~Girrbach, {\it {Left-handed $Z'$ and $Z$ FCNC quark
  couplings facing new $b \to s \mu^+ \mu^-$ data}},  {\em JHEP} {\bf 12}
  (2013) 009, [\href{http://arxiv.org/abs/1309.2466}{{\tt arXiv:1309.2466}}].

\bibitem{Altmannshofer:2014cfa}
W.~Altmannshofer, S.~Gori, M.~Pospelov, and I.~Yavin, {\it {Quark flavor
  transitions in $L_\mu-L_\tau$ models}},  {\em Phys. Rev.} {\bf D89} (2014)
  095033, [\href{http://arxiv.org/abs/1403.1269}{{\tt arXiv:1403.1269}}].

\bibitem{Crivellin:2015mga}
A.~Crivellin, G.~D'Ambrosio, and J.~Heeck, {\it {Explaining
  $h\to\mu^\pm\tau^\mp$, $B\to K^* \mu^+\mu^-$ and $B\to K \mu^+\mu^-/B\to K
  e^+e^-$ in a two-Higgs-doublet model with gauged $L_\mu-L_\tau$}},  {\em
  Phys. Rev. Lett.} {\bf 114} (2015) 151801,
  [\href{http://arxiv.org/abs/1501.00993}{{\tt arXiv:1501.00993}}].

\bibitem{Crivellin:2015lwa}
A.~Crivellin, G.~D'Ambrosio, and J.~Heeck, {\it {Addressing the LHC flavor
  anomalies with horizontal gauge symmetries}},  {\em Phys. Rev.} {\bf D91}
  (2015), no.~7 075006, [\href{http://arxiv.org/abs/1503.03477}{{\tt
  arXiv:1503.03477}}].

\bibitem{Niehoff:2015bfa}
C.~Niehoff, P.~Stangl, and D.~M. Straub, {\it {Violation of lepton flavour
  universality in composite Higgs models}},  {\em Phys. Lett.} {\bf B747}
  (2015) 182--186, [\href{http://arxiv.org/abs/1503.03865}{{\tt
  arXiv:1503.03865}}].

\bibitem{Celis:2015ara}
A.~Celis, J.~Fuentes-Martin, M.~Jung, and H.~Serodio, {\it {Family nonuniversal
  $Z\prime$ models with protected flavor-changing interactions}},  {\em Phys.
  Rev.} {\bf D92} (2015), no.~1 015007,
  [\href{http://arxiv.org/abs/1505.03079}{{\tt arXiv:1505.03079}}].

\bibitem{Greljo:2015mma}
A.~Greljo, G.~Isidori, and D.~Marzocca, {\it {On the breaking of Lepton Flavor
  Universality in B decays}},  {\em JHEP} {\bf 07} (2015) 142,
  [\href{http://arxiv.org/abs/1506.01705}{{\tt arXiv:1506.01705}}].

\bibitem{Niehoff:2015iaa}
C.~Niehoff, P.~Stangl, and D.~M. Straub, {\it {Direct and indirect signals of
  natural composite Higgs models}},  {\em JHEP} {\bf 01} (2016) 119,
  [\href{http://arxiv.org/abs/1508.00569}{{\tt arXiv:1508.00569}}].

\bibitem{Altmannshofer:2015mqa}
W.~Altmannshofer and I.~Yavin, {\it {Predictions for lepton flavor universality
  violation in rare B decays in models with gauged $L_\mu - L_\tau$}},  {\em
  Phys. Rev.} {\bf D92} (2015), no.~7 075022,
  [\href{http://arxiv.org/abs/1508.07009}{{\tt arXiv:1508.07009}}].

\bibitem{Falkowski:2015zwa}
A.~Falkowski, M.~Nardecchia, and R.~Ziegler, {\it {Lepton Flavor
  Non-Universality in B-meson Decays from a U(2) Flavor Model}},  {\em JHEP}
  {\bf 11} (2015) 173, [\href{http://arxiv.org/abs/1509.01249}{{\tt
  arXiv:1509.01249}}].

\bibitem{Carmona:2015ena}
A.~Carmona and F.~Goertz, {\it {Lepton Flavor and Nonuniversality from Minimal
  Composite Higgs Setups}},  {\em Phys. Rev. Lett.} {\bf 116} (2016), no.~25
  251801, [\href{http://arxiv.org/abs/1510.07658}{{\tt arXiv:1510.07658}}].

\bibitem{GarciaGarcia:2016nvr}
I.~Garcia~Garcia, {\it {LHCb anomalies from a natural perspective}},  {\em
  JHEP} {\bf 03} (2017) 040, [\href{http://arxiv.org/abs/1611.03507}{{\tt
  arXiv:1611.03507}}].

\bibitem{Megias:2016bde}
E.~Megias, G.~Panico, O.~Pujolas, and M.~Quiros, {\it {A Natural origin for the
  LHCb anomalies}},  {\em JHEP} {\bf 09} (2016) 118,
  [\href{http://arxiv.org/abs/1608.02362}{{\tt arXiv:1608.02362}}].

\bibitem{Chiang:2016qov}
C.-W. Chiang, X.-G. He, and G.~Valencia, {\it {$Z\prime$ model for $b\to s \ell
  \overline{\ell}$ flavor anomalies}},  {\em Phys. Rev.} {\bf D93} (2016),
  no.~7 074003, [\href{http://arxiv.org/abs/1601.07328}{{\tt
  arXiv:1601.07328}}].

\bibitem{Altmannshofer:2016oaq}
W.~Altmannshofer, M.~Carena, and A.~Crivellin, {\it {$L_\mu - L_\tau$ theory of
  Higgs flavor violation and $(g-2)_\mu$}},  {\em Phys. Rev.} {\bf D94} (2016),
  no.~9 095026, [\href{http://arxiv.org/abs/1604.08221}{{\tt
  arXiv:1604.08221}}].

\bibitem{Boucenna:2016qad}
S.~M. Boucenna, A.~Celis, J.~Fuentes-Martin, A.~Vicente, and J.~Virto, {\it
  {Phenomenology of an $SU(2) \times SU(2) \times U(1)$ model with
  lepton-flavour non-universality}},  {\em JHEP} {\bf 12} (2016) 059,
  [\href{http://arxiv.org/abs/1608.01349}{{\tt arXiv:1608.01349}}].

\bibitem{Foldenauer:2016rpi}
P.~Foldenauer and J.~Jaeckel, {\it {Purely flavor-changing Z' bosons and where
  they might hide}},  {\em JHEP} {\bf 05} (2017) 010,
  [\href{http://arxiv.org/abs/1612.07789}{{\tt arXiv:1612.07789}}].

\bibitem{Kamenik:2017tnu}
J.~F. Kamenik, Y.~Soreq, and J.~Zupan, {\it {Lepton flavor universality
  violation without new sources of quark flavor violation}},  {\em Phys. Rev.}
  {\bf D97} (2018), no.~3 035002, [\href{http://arxiv.org/abs/1704.06005}{{\tt
  arXiv:1704.06005}}].

\bibitem{Chivukula:2017qsi}
R.~S. Chivukula, J.~Isaacson, K.~A. Mohan, D.~Sengupta, and E.~H. Simmons, {\it
  {$R_K$ anomalies and simplified limits on $Z'$ models at the LHC}},  {\em
  Phys. Rev.} {\bf D96} (2017), no.~7 075012,
  [\href{http://arxiv.org/abs/1706.06575}{{\tt arXiv:1706.06575}}].

\bibitem{Faisel:2017glo}
G.~Faisel and J.~Tandean, {\it {Connecting $ b\to s\ell \overline{\ell} $
  anomalies to enhanced rare nonleptonic ${\overline{B}}_s^0$ decays in
  $Z^\prime$ model}},  {\em JHEP} {\bf 02} (2018) 074,
  [\href{http://arxiv.org/abs/1710.11102}{{\tt arXiv:1710.11102}}].

\bibitem{Ellis:2017nrp}
J.~Ellis, M.~Fairbairn, and P.~Tunney, {\it {Anomaly-Free Models for Flavour
  Anomalies}},  \href{http://arxiv.org/abs/1705.03447}{{\tt arXiv:1705.03447}}.

\bibitem{Alonso:2017uky}
R.~Alonso, P.~Cox, C.~Han, and T.~T. Yanagida, {\it {Flavoured $B-L$ local
  symmetry and anomalous rare $B$ decays}},  {\em Phys. Lett.} {\bf B774}
  (2017) 643--648, [\href{http://arxiv.org/abs/1705.03858}{{\tt
  arXiv:1705.03858}}].

\bibitem{Carmona:2017fsn}
A.~Carmona and F.~Goertz, {\it {Recent $\boldsymbol{B}$ Physics Anomalies - a
  First Hint for Compositeness?}},  \href{http://arxiv.org/abs/1712.02536}{{\tt
  arXiv:1712.02536}}.

\bibitem{Dalchenko:2017shg}
M.~Abdullah, M.~Dalchenko, B.~Dutta, R.~Eusebi, P.~Huang, T.~Kamon,
  D.~Rathjens, and A.~Thompson, {\it {Bottom-quark fusion processes at the LHC
  for probing $Z'$ models and $B$-meson decay anomalies}},  {\em Phys. Rev.}
  {\bf D97} (2018), no.~7 075035, [\href{http://arxiv.org/abs/1707.07016}{{\tt
  arXiv:1707.07016}}].

\bibitem{Raby:2017igl}
S.~Raby and A.~Trautner, {\it {A "Vector-like chiral" fourth family to explain
  muon anomalies}},  \href{http://arxiv.org/abs/1712.09360}{{\tt
  arXiv:1712.09360}}.

\bibitem{Bian:2017rpg}
L.~Bian, S.-M. Choi, Y.-J. Kang, and H.~M. Lee, {\it {A minimal flavored
  $U(1)'$ for $B$-meson anomalies}},  {\em Phys. Rev.} {\bf D96} (2017), no.~7
  075038, [\href{http://arxiv.org/abs/1707.04811}{{\tt arXiv:1707.04811}}].

\bibitem{Bian:2017xzg}
L.~Bian, H.~M. Lee, and C.~B. Park, {\it {$B$-meson anomalies and Higgs physics
  in flavored $U(1)'$ model}},  \href{http://arxiv.org/abs/1711.08930}{{\tt
  arXiv:1711.08930}}.

\bibitem{Alok:2017jgr}
A.~K. Alok, B.~Bhattacharya, D.~Kumar, J.~Kumar, D.~London, and S.~U. Sankar,
  {\it {New physics in $b \rightarrow s \mu^+ \mu^-$: Distinguishing models
  through CP-violating effects}},  {\em Phys. Rev.} {\bf D96} (2017), no.~1
  015034, [\href{http://arxiv.org/abs/1703.09247}{{\tt arXiv:1703.09247}}].

\bibitem{Falkowski:2018dsl}
A.~Falkowski, S.~F. King, E.~Perdomo, and M.~Pierre, {\it {Flavourful $Z'$
  portal for vector-like neutrino Dark Matter and $R_{K^{(*)}}$}},  {\em JHEP}
  {\bf 08} (2018) 061, [\href{http://arxiv.org/abs/1803.04430}{{\tt
  arXiv:1803.04430}}].

\bibitem{Kohda:2018xbc}
M.~Kohda, T.~Modak, and A.~Soffer, {\it {Identifying a $Z'$ behind $b \to s
  \ell \ell$ anomalies at the LHC}},  {\em Phys. Rev.} {\bf D97} (2018), no.~11
  115019, [\href{http://arxiv.org/abs/1803.07492}{{\tt arXiv:1803.07492}}].

\bibitem{Fox:2018ldq}
P.~J. Fox, I.~Low, and Y.~Zhang, {\it {Top-philic $Z'$ Forces at the LHC}},
  \href{http://arxiv.org/abs/1801.03505}{{\tt arXiv:1801.03505}}.

\bibitem{Chala:2018igk}
M.~Chala and M.~Spannowsky, {\it {On the behaviour of composite resonances
  breaking lepton flavour universality}},
  \href{http://arxiv.org/abs/1803.02364}{{\tt arXiv:1803.02364}}.

\bibitem{Darme:2018hqg}
L.~Darmé, K.~Kowalska, L.~Roszkowski, and E.~M. Sessolo, {\it {Flavor
  anomalies and dark matter in SUSY with an extra U(1)}},  {\em JHEP} {\bf 10}
  (2018) 052, [\href{http://arxiv.org/abs/1806.06036}{{\tt arXiv:1806.06036}}].

\bibitem{Allanach:2018odd}
B.~C. Allanach, T.~Corbett, M.~J. Dolan, and T.~You, {\it {Hadron collider
  sensitivity to fat flavourful Z' for ${R}_{K^{\left(\ast \right)}}$}},  {\em
  JHEP} {\bf 03} (2019) 137, [\href{http://arxiv.org/abs/1810.02166}{{\tt
  arXiv:1810.02166}}].

\bibitem{Ko:2019tts}
P.~Ko, T.~Nomura, and C.~Yu, {\it {$b\rightarrow s \mu^+ \mu^-$ anomalies and
  related phenomenology in $U(1)_{B_3 - x_\mu L_\mu - x_\tau L_\tau}$ flavor
  gauge models}},  {\em JHEP} {\bf 04} (2019) 102,
  [\href{http://arxiv.org/abs/1902.06107}{{\tt arXiv:1902.06107}}].

\bibitem{Biswas:2019twf}
A.~Biswas and A.~Shaw, {\it {Reconciling dark matter, $R_{K^{(*)}}$ anomalies
  and $(g-2)_{\mu}$ in an ${L_{\mu}-L_{\tau}}$ scenario}},  {\em JHEP} {\bf 05}
  (2019) 165, [\href{http://arxiv.org/abs/1903.08745}{{\tt arXiv:1903.08745}}].

\bibitem{Allanach:2019mfl}
B.~C. Allanach, J.~M. Butterworth, and T.~Corbett, {\it {Collider Constraints
  on $Z^\prime$ Models for Neutral Current $B-$Anomalies}},
  \href{http://arxiv.org/abs/1904.10954}{{\tt arXiv:1904.10954}}.

\bibitem{Alok:2019ufo}
A.~K. Alok, A.~Dighe, S.~Gangal, and D.~Kumar, {\it {Continuing search for new
  physics in $b \to s \mu \mu$ decays: two operators at a time}},  {\em JHEP}
  {\bf 06} (2019) 089, [\href{http://arxiv.org/abs/1903.09617}{{\tt
  arXiv:1903.09617}}].

\bibitem{Kawamura:2019rth}
J.~Kawamura, S.~Raby, and A.~Trautner, {\it {Complete Vector-like Fourth Family
  and new $\mathrm{U(1)^\prime}$ for Muon Anomalies}},
  \href{http://arxiv.org/abs/1906.11297}{{\tt arXiv:1906.11297}}.

\bibitem{Geiregat:1990gz}
{\bf CHARM-II} Collaboration, D.~Geiregat et~al., {\it {First observation of
  neutrino trident production}},  {\em Phys. Lett.} {\bf B245} (1990) 271--275.

\bibitem{Mishra:1991bv}
{\bf CCFR} Collaboration, S.~R. Mishra et~al., {\it {Neutrino tridents and W Z
  interference}},  {\em Phys. Rev. Lett.} {\bf 66} (1991) 3117--3120.

\bibitem{Altmannshofer:2014pba}
W.~Altmannshofer, S.~Gori, M.~Pospelov, and I.~Yavin, {\it {Neutrino Trident
  Production: A Powerful Probe of New Physics with Neutrino Beams}},  {\em
  Phys. Rev. Lett.} {\bf 113} (2014) 091801,
  [\href{http://arxiv.org/abs/1406.2332}{{\tt arXiv:1406.2332}}].

\bibitem{Falkowski:2017pss}
A.~Falkowski, M.~González-Alonso, and K.~Mimouni, {\it {Compilation of
  low-energy constraints on 4-fermion operators in the SMEFT}},  {\em JHEP}
  {\bf 08} (2017) 123, [\href{http://arxiv.org/abs/1706.03783}{{\tt
  arXiv:1706.03783}}].

\bibitem{Charles:2004jd}
{\bf CKMfitter Group} Collaboration, J.~Charles, A.~Hocker, H.~Lacker,
  S.~Laplace, F.~R. Le~Diberder, J.~Malcles, J.~Ocariz, M.~Pivk, and L.~Roos,
  {\it {CP violation and the CKM matrix: Assessing the impact of the asymmetric
  $B$ factories}},  {\em Eur. Phys. J.} {\bf C41} (2005), no.~1 1--131,
  [\href{http://arxiv.org/abs/hep-ph/0406184}{{\tt hep-ph/0406184}}].

\bibitem{Descotes-Genon:2018foz}
S.~Descotes-Genon, A.~Falkowski, M.~Fedele, M.~González-Alonso, and J.~Virto,
  {\it {The CKM parameters in the SMEFT}},  {\em JHEP} {\bf 05} (2019) 172,
  [\href{http://arxiv.org/abs/1812.08163}{{\tt arXiv:1812.08163}}].

\bibitem{Altmannshofer:2016jzy}
W.~Altmannshofer, S.~Gori, S.~Profumo, and F.~S. Queiroz, {\it {Explaining dark
  matter and B decay anomalies with an $L_\mu - L_\tau$ model}},  {\em JHEP}
  {\bf 12} (2016) 106, [\href{http://arxiv.org/abs/1609.04026}{{\tt
  arXiv:1609.04026}}].

\bibitem{Alloul:2013bka}
A.~Alloul, N.~D. Christensen, C.~Degrande, C.~Duhr, and B.~Fuks, {\it
  {FeynRules 2.0 - A complete toolbox for tree-level phenomenology}},  {\em
  Comput. Phys. Commun.} {\bf 185} (2014) 2250--2300,
  [\href{http://arxiv.org/abs/1310.1921}{{\tt arXiv:1310.1921}}].

\bibitem{Alwall:2014hca}
J.~Alwall, R.~Frederix, S.~Frixione, V.~Hirschi, F.~Maltoni, O.~Mattelaer,
  H.~S. Shao, T.~Stelzer, P.~Torrielli, and M.~Zaro, {\it {The automated
  computation of tree-level and next-to-leading order differential cross
  sections, and their matching to parton shower simulations}},  {\em JHEP} {\bf
  07} (2014) 079, [\href{http://arxiv.org/abs/1405.0301}{{\tt
  arXiv:1405.0301}}].

\bibitem{Ball:2014uwa}
{\bf NNPDF} Collaboration, R.~D. Ball et~al., {\it {Parton distributions for
  the LHC Run II}},  {\em JHEP} {\bf 04} (2015) 040,
  [\href{http://arxiv.org/abs/1410.8849}{{\tt arXiv:1410.8849}}].

\bibitem{Sjostrand:2014zea}
T.~Sjostrand, S.~Ask, J.~R. Christiansen, R.~Corke, N.~Desai, P.~Ilten,
  S.~Mrenna, S.~Prestel, C.~O. Rasmussen, and P.~Z. Skands, {\it {An
  Introduction to PYTHIA 8.2}},  {\em Comput. Phys. Commun.} {\bf 191} (2015)
  159--177, [\href{http://arxiv.org/abs/1410.3012}{{\tt arXiv:1410.3012}}].

\bibitem{deFavereau:2013fsa}
{\bf DELPHES 3} Collaboration, J.~de~Favereau, C.~Delaere, P.~Demin,
  A.~Giammanco, V.~LemaÃ®tre, A.~Mertens, and M.~Selvaggi, {\it {DELPHES 3, A
  modular framework for fast simulation of a generic collider experiment}},
  {\em JHEP} {\bf 02} (2014) 057, [\href{http://arxiv.org/abs/1307.6346}{{\tt
  arXiv:1307.6346}}].

\bibitem{Cacciari:2008gp}
M.~Cacciari, G.~P. Salam, and G.~Soyez, {\it {The anti-$k_t$ jet clustering
  algorithm}},  {\em JHEP} {\bf 04} (2008) 063,
  [\href{http://arxiv.org/abs/0802.1189}{{\tt arXiv:0802.1189}}].

\bibitem{Mangano:2006rw}
M.~L. Mangano, M.~Moretti, F.~Piccinini, and M.~Treccani, {\it {Matching matrix
  elements and shower evolution for top-quark production in hadronic
  collisions}},  {\em JHEP} {\bf 01} (2007) 013,
  [\href{http://arxiv.org/abs/hep-ph/0611129}{{\tt hep-ph/0611129}}].

\bibitem{Hoche:2006ph}
S.~Hoeche, F.~Krauss, N.~Lavesson, L.~Lonnblad, M.~Mangano, A.~Schalicke, and
  S.~Schumann, {\it {Matching parton showers and matrix elements}},  in {\em
  {HERA and the LHC: A Workshop on the implications of HERA for LHC physics:
  Proceedings Part A}}, pp.~288--289, 2005.
\newblock \href{http://arxiv.org/abs/hep-ph/0602031}{{\tt hep-ph/0602031}}.

\bibitem{Chatrchyan:2012jua}
{\bf CMS} Collaboration, S.~Chatrchyan et~al., {\it {Identification of b-quark
  jets with the CMS experiment}},  {\em JINST} {\bf 8} (2013) P04013,
  [\href{http://arxiv.org/abs/1211.4462}{{\tt arXiv:1211.4462}}].

\bibitem{Cowan:2010js}
G.~Cowan, K.~Cranmer, E.~Gross, and O.~Vitells, {\it {Asymptotic formulae for
  likelihood-based tests of new physics}},  {\em Eur. Phys. J.} {\bf C71}
  (2011) 1554, [\href{http://arxiv.org/abs/1007.1727}{{\tt arXiv:1007.1727}}].
  [Erratum: Eur. Phys. J.C73,2501(2013)].

\bibitem{Campbell:2005zv}
J.~M. Campbell, R.~K. Ellis, F.~Maltoni, and S.~Willenbrock, {\it {Production
  of a $Z$ boson and two jets with one heavy-quark tag}},  {\em Phys. Rev.}
  {\bf D73} (2006) 054007, [\href{http://arxiv.org/abs/hep-ph/0510362}{{\tt
  hep-ph/0510362}}]. [Erratum: Phys. Rev.D77,019903(2008)].

\bibitem{Bern:2008ef}
{\bf NLO Multileg Working Group} Collaboration, Z.~Bern et~al., {\it {The NLO
  multileg working group: Summary report}},  in {\em {Physics at TeV colliders,
  La physique du TeV aux collisionneurs, Les Houches 2007 : 11-29 June 2007}},
  pp.~1--120, 2008.
\newblock \href{http://arxiv.org/abs/0803.0494}{{\tt arXiv:0803.0494}}.

\bibitem{Catani:2009sm}
S.~Catani, L.~Cieri, G.~Ferrera, D.~de~Florian, and M.~Grazzini, {\it {Vector
  boson production at hadron colliders: a fully exclusive QCD calculation at
  NNLO}},  {\em Phys. Rev. Lett.} {\bf 103} (2009) 082001,
  [\href{http://arxiv.org/abs/0903.2120}{{\tt arXiv:0903.2120}}].

\bibitem{Chatrchyan:2014aea}
{\bf CMS} Collaboration, S.~Chatrchyan et~al., {\it {Search for anomalous
  production of events with three or more leptons in $pp$ collisions at
  $\sqrt(s) =$ 8 TeV}},  {\em Phys. Rev.} {\bf D90} (2014) 032006,
  [\href{http://arxiv.org/abs/1404.5801}{{\tt arXiv:1404.5801}}].

\bibitem{Sirunyan:2017lae}
{\bf CMS} Collaboration, A.~M. Sirunyan et~al., {\it {Search for electroweak
  production of charginos and neutralinos in multilepton final states in
  proton-proton collisions at $\sqrt{s}=$ 13 TeV}},  {\em JHEP} {\bf 03} (2018)
  166, [\href{http://arxiv.org/abs/1709.05406}{{\tt arXiv:1709.05406}}].

\bibitem{ATLAS:2016pbt}
{\bf ATLAS} Collaboration, T.~A. collaboration, {\it {Search for doubly-charged
  Higgs bosons in same-charge electron pair final states using proton-proton
  collisions at $\sqrt{s}=13\,\mathrm{TeV}$ with the ATLAS detector,
  ATLAS-CONF-2016-051}}, .

\bibitem{Grazzini:2016swo}
M.~Grazzini, S.~Kallweit, D.~Rathlev, and M.~Wiesemann, {\it {$W^{\pm}Z$
  production at hadron colliders in NNLO QCD}},  {\em Phys. Lett.} {\bf B761}
  (2016) 179--183, [\href{http://arxiv.org/abs/1604.08576}{{\tt
  arXiv:1604.08576}}].

\bibitem{Aaboud:2017yvp}
{\bf ATLAS} Collaboration, M.~Aaboud et~al., {\it {Search for new phenomena in
  dijet events using 37 fb$^{-1}$ of $pp$ collision data collected at
  $\sqrt{s}=$13 TeV with the ATLAS detector}},  {\em Phys. Rev.} {\bf D96}
  (2017), no.~5 052004, [\href{http://arxiv.org/abs/1703.09127}{{\tt
  arXiv:1703.09127}}].

\bibitem{CMS:2017xrr}
{\bf CMS} Collaboration, C.~Collaboration, {\it {Searches for dijet resonances
  in pp collisions at $\sqrt{s}=13~\mathrm{TeV}$ using data collected in
  2016,CMS-PAS-EXO-16-056}}, .

\bibitem{ATLAS:2019vcr}
{\bf ATLAS} Collaboration, T.~A. collaboration, {\it {Search for high-mass
  dilepton resonances using $139\,\mathrm{fb}^{-1}$ of $pp$ collision data
  collected at $\sqrt{s}=13\,\mathrm{TeV}$ with the ATLAS
  detector,ATLAS-CONF-2019-001}}, .

\bibitem{Aad:2019fac}
{\bf ATLAS} Collaboration, G.~Aad et~al., {\it {Search for high-mass dilepton
  resonances using 139 fb$^{-1}$ of $pp$ collision data collected at
  $\sqrt{s}=$13 TeV with the ATLAS detector}},
  \href{http://arxiv.org/abs/1903.06248}{{\tt arXiv:1903.06248}}.

\end{thebibliography}\endgroup

\end{document}